# Review of the high-power vacuum tube microwave sources

Weiye Xu [1,a)]

[1]Institute of Plasma Physics, Chinese Academy of Sciences, 230031 Hefei, Anhui, China

[a)]Electronic address: xuweiye@ipp.cas.cn

**Data availability statement:** The data that supports the findings of this study are available within the article.

## Abstract

Since the first vacuum tube (X-ray tube) was invented by Wilhelm Röntgen in Germany, after more than one hundred years of development, the average power density of the vacuum tube microwave source has reached the order of $10^8$ [MW][GHz]$^2$. The maximum power density record was created by the Free Electron Lasers. In the high-power microwave field, the vacuum devices are still the mainstream microwave sources for applications such as scientific instruments, communications, radars, magnetic confinement fusion heating, microwave weapons, etc. The principles of microwave generation by vacuum tube microwave sources include Cherenkov or Smith-Purcell radiation, transition radiation, and Bremsstrahlung. In this paper, the vacuum tube microwave sources were reviewed in order according to the three radiation principles. Among them, the Vircators can produce 22 GW output power in P-band (0.23-1GHz). Vacuum tubes that can achieve continuous-wave operation include Traveling Wave Tubes, Klystrons, Free Electron Lasers, and Magnetrons, with output power up to 1MW. Vacuum tubes that can generate frequencies of the order of 100 GHz and above include Klystrons, Traveling Wave Tubes, Gyrotrons, Backward Wave Oscillators, Magnetrons, Surface Wave Oscillators, Free Electron Lasers, Orotrons, etc. Gyrotrons are very attractive in the millimeter wave and THz fields. The Gyrotrons can output power at the MW level with 3000s pulse width at millimeter wavelengths.

# 1. Introduction

High-power microwave (HPM) source generally refers to a microwave source with a peak output power exceeding 100MW or an average output power exceeding 100kW and a frequency range between 0.3-300GHz. The performance index for evaluating high-power microwave sources is the quality factor [1] $P_{av}f^2$ ($P_{av}$ is the average power, and $f$ is the frequency). The higher the quality factor, the better. Driven by applications such as magnetically confinement nuclear fusion, microwave-assisted drilling, microwave weapons, communications, radar, high-energy RF accelerators, wireless power transmission, and material processing, the quality factor of microwave devices is increasing at an order of magnitude every ten years. Looking at the actual demand, HPM technology is developing towards the goals of high power, high efficiency, and wide frequency band. In terms of power, there are two development directions, one is high peak power and the other is high average power.

High-power microwave sources can be divided into two categories, pulse microwave sources and non-pulse microwave sources. Generally speaking, the pulse microwave source refers to a microwave source whose rising edge is on the order of sub-nanoseconds or picoseconds. It includes a primary drive (pulse generation system or explosive), a pulse compression system, a microwave source, and an antenna. Pulse sources usually convert energy into short-pulse electromagnetic radiation through low-speed storage and rapid release of energy. The main technologies for pulse generation include Marx generators [2,3], Blumlein Line [4,5] technology, etc. In electromagnetic bombs, explosives are used as the primary drive, and the shock wave generated by the explosion drives a pulse compressor to generate a strong electron flow. Among various vacuum tube microwave sources, the virtual cathode oscillator (VCO) is the most popular choice for making pulse sources such as microwave bombs. It has a simple and compact structure and can generate a strong wide-spectrum single-pulse microwave. There are also many reports of using backward wave oscillators (BWO) as the pulse microwave sources [6]. In addition, there are ultra-wideband pulse microwave sources that feed high-power pulses directly to the antenna without using a vacuum tube [7,8]. The non-pulse microwave source uses a continuously operating high-voltage power source or a pulsed high-voltage power source with a relatively high duty cycle to drive the electron gun to generate an electron beam.

In 1896, Wilhelm Röntgen invented the first vacuum tube in Germany, the X-ray tube. In 1904, the British physicist John Ambrose Fleming invented the first vacuum diode. In 1906, Lee de Forest invented the first vacuum triode in the United States. Gradually, vacuum tubes have been used in more and more applications, including control devices, scientific instruments, electronic gramophones, FM radios, televisions, radars, sonars, etc. With the invention and development of solid-state devices, the application of vacuum devices in the fields of communications and consumer electronics has gradually withdrawn from the historical stage. But in the high-power microwave field, vacuum devices are still the mainstream.

The magnetron was granted its first patent in 1935, and in 1940 the British first deployed the magnetron on a radar. In 1937, the first cavity-type device - klystron was born. After decades of development, many types of high-power vacuum tube microwave sources have been produced. Since 1960, the development of high-energy physics theory and technology has promoted the introduction of pulsed power technology. The generation of high-current (I> MA) relativistic electron beams with energy close to the static energy of electrons (510 keV) and high-voltage pulses with voltages of several megavolts or higher has become a reality, which has expanded the range of high-power microwaves. A large number of high-voltage operating devices that rely on strong currents, such as relativistic klystron and virtual cathode oscillator (VCO), have emerged. At the same time, some devices based on the relativity effect, such as gyrotron and free electron laser (FEL), have also appeared. Table I lists the current main high-power microwave sources, which can be divided into three types: O-type, M-type, and space-charge type.

O-type device refers to the device whose electron beam drifts in the same direction as the applied magnetic field. Among them, O-type slow-wave devices use the axial slow-wave structure to achieve electron beam clustering and beam interaction. The frequency is stable, the beam interaction efficiency is high, but its high impedance property limits the generation and improvement of power, and the high magnetic field limits the miniaturization of the device.

M-type device refers to the device whose drift direction of the relativistic electron beam is perpendicular to the magnetic field. The average drift velocity ($v_d = E/B$) of the electron under the action of the electromagnetic field is equal to the phase velocity $v_p$ of the RF wave. The M-type microwave tube has a compact structure, a low working voltage, and a high efficiency, which can reach more than 80%. Such microwave tubes include magnetrons and crossed-field amplifiers (CFAs), which can be used in radar transmitters, electronic countermeasure technology, linear accelerators, microwave heating and other fields.

High-power microwave devices with space charge effects, such as virtual cathode oscillators [9], generally do not require an external guidance magnetic field for operation. Compared with other types of high-power microwave sources, they have the advantages of simple structure, low requirements on the quality of the electron beam, high power capacity, relatively easy tuning, low impedance, etc. However, they also have shortcomings such as relatively low beam-wave power conversion efficiency, a messy frequency spectrum, and impure modes.

TABLE I. High-power vacuum tube microwave sources. The red (light gray shading) in the table indicates the strong current relativity devices, the purple (dark gray shading) indicates the weak current relativity devices, and the black (no shading) indicates the non-relativistic devices.

| | **Slow wave ($v_p < c$)** | **Fast wave ($v_p > c$)** |
|---|---|---|
| **O-type** | Traveling Wave Tube (TWT) | Free Electron Laser (FEL) |
| | Relativistic TWT | Gyrotron |
| | Backward Wave Oscillator (BWO) | Gyro BWO |
| | Relativistic BWO | Gyro TWT |
| | Multi-Save Cherenkov Generator (MWCG) | Gyro Klystron |
| | Relativistic Diffraction Generator (RDG) | Gyro Twystron |
| | Surface Save Oscillator (SWO) | Cyclotron Auto-Resonance Maser (CARM) |
| | Orotron | Magnicon |
| | Flimatron | |
| | Klystron | |
| | Relativistic Klystron | |
| | Transit Time Oscillator (TTO) | |
| **M-type** | Magnetron | |
| | Relativistic Magnetron | |
| | Crossed-Field Amplifier (CFA) | |
| | Magnetically Insulated transmission Line Oscillator (MILO) | |
| **Space-charge type** | Virtual Cathode Oscillator (VCO) | |

Vacuum tube microwave sources can be divided into high-current relativity devices, weak-current relativity devices, and non-relativity devices according to the properties of the electron beam. High-current relativistic beam microwave sources are used to generate short-pulse, high-peak-power microwaves and are used in fields such as accelerators. At present, the peak power of several typical high-current relativistic microwave devices has reached the level of GW to dozens of GW, the pulse width can reach 100 ns, and the frequency ranges from P band (230MHz ~ 1000MHz) to millimeter wave band, and the energy of a single pulse is tens to thousands of joules, with a repetition frequency of tens of Hz. Weak-current relativistic and non-relativistic microwave sources can generate microwaves with high average power and long pulses. They are the microwave sources required for plasma heating in magnetic confinement fusion.

The main principles of microwave generation by vacuum tube microwave sources are as follows: slow-wave Cherenkov or Smith-Purcell radiation, transition radiation, and Bremsstrahlung [1,10].

Cherenkov radiation refers to the radiation generated when electrons move in a medium when the speed of the electrons is greater than the speed of the electromagnetic waves moving in the medium. It also includes radiation when the electrons move in a periodic slow wave structure. Typical devices based on Cherenkov radiation include traveling wave tubes (TWTs), backward wave oscillators (BWOs), surface wave oscillators (SWOs), multi-wave Cherenkov generators (MWCGs), relativistic diffraction generators (RDGs), Magnetrons, Crossed-Field Amplifier (CFA), magnetically insulated transmission line oscillators (MILOs), etc.

Transition radiation refers to the radiation when electrons pass through interfaces with different refractive indices, and also includes radiation when passing through disturbances in the same medium such as conductive grids, metal sheets or gaps on the surface of conductors. The main difference from Cherenkov radiation is that the field interacting with the electron beam is a standing wave field. Devices based on the transit radiation include transit time oscillators (TTOs), klystrons, and the like.

Bremsstrahlung refers to the radiation when electrons move in an external electromagnetic field at a varying speed. Generally speaking, electrons move in an oscillating form. At this time, the frequency of the electromagnetic wave radiated by the electron is consistent with the frequency of its oscillation, or the frequency of a certain harmonic that it oscillates. Microwave sources based on this include free electron lasers (FELs), gyrotrons, and Cyclotron Auto-Resonance Masers (CARMs).

Theories for analyzing the beam-wave interaction of vacuum tube microwave sources include small-signal linear theory and large-signal nonlinear theory. The beam-wave interactions in HPM devices are complex, making accurate analytical analysis difficult, and the cost of experimenting and developing high-power microwave sources is high. In order to accurately analyze the beam interaction state, a large number of numerical simulation methods have been studied, and the PIC (Particle In Cell) method has been widely used. John M. Dawson summarized the PIC method in 1983 [11]. The PIC method is based on the concept of "macro particles" [12]. PIC-based software includes MAGIC [13] developed by MRC (Mission Research Corporation), Germany's CST MAFIA (now integrated into the CST Particle Studio), XOOPIC, MICHELLE developed by the University of California, Berkeley, and so on. China's Northwest Institute of Nuclear Technology and Xi'an Jiaotong University have jointly developed a 2.5-dimensional PIC simulation software UNIPIC [14], which can simulate high-power microwave devices such as magnetron, VCO, BWO, and MILO.

Electron-optical systems involve the generation, shaping, maintenance and collection of electron beams. At present, many electron-optical system design simulation programs have been developed, including EGUN, CAMEO, etc. EGUN is developed by Bill Herrmannsfeldt at Stanford Linear Accelerator Center (SLAC). It is powerful for electronic trajectory calculation [15, 16]; CAMEO (CAMbridge Electron Optics) is an electron gun design software designed by Cambridge University.

Compared to vacuum devices, the development of solid-state devices is more widely known. Since Intel released the world's first microprocessor 4004 in 1971, the density of electronic components integrated on the chip has increased by seven orders of magnitude, and its growth rate follows Moore's law and doubles every two years. In fact, the development of vacuum devices also follows Moore's Law, but it is not the density of electronic components that increases, but the average power density ([MW] [GHz]$^2$). The average power density progress of the main types of vacuum tubes in the 20th century [17] is shown in Fig. 1.

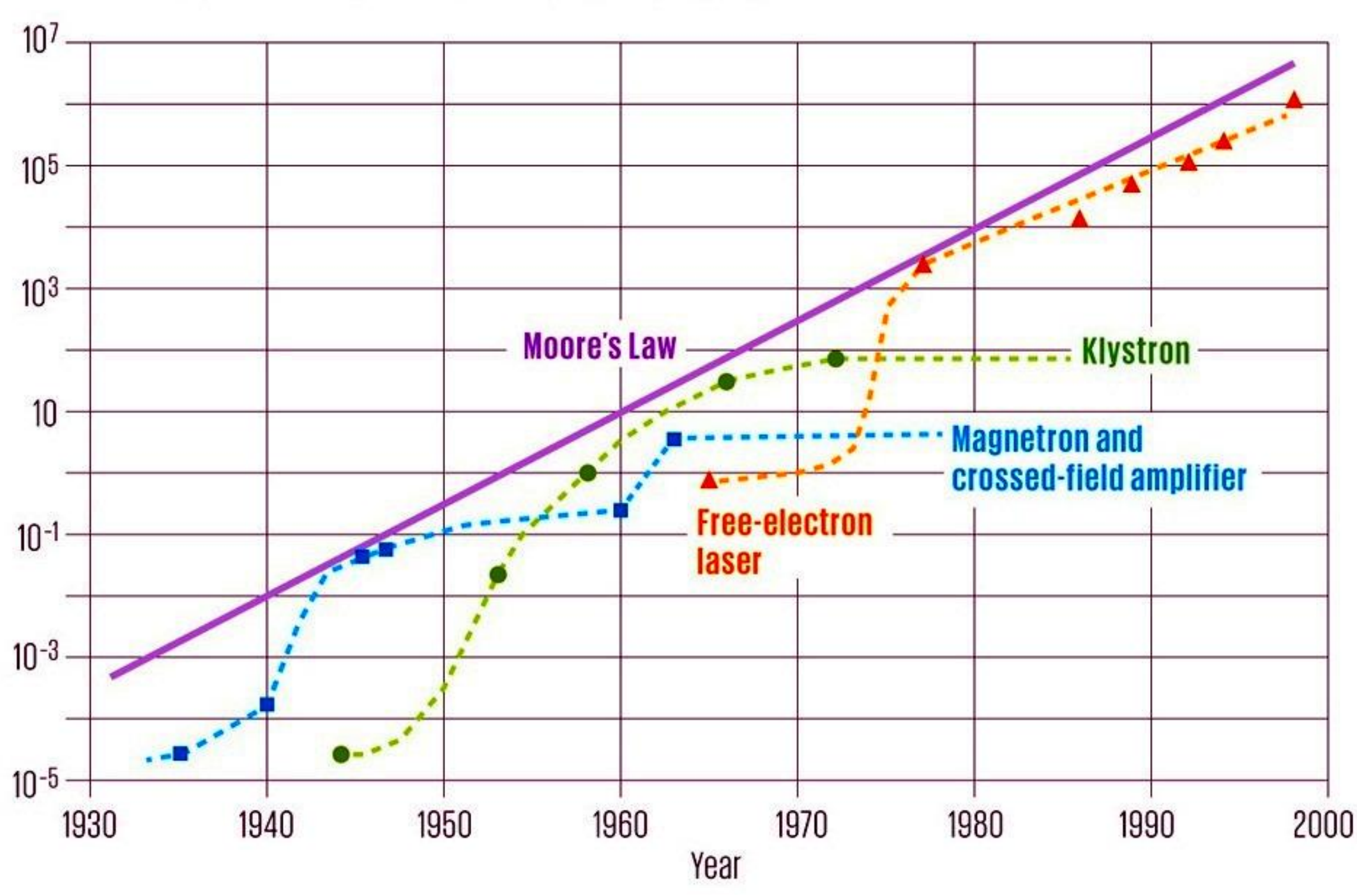

FIG. 1. Progress in the power density of the main types of vacuum tubes in the 20th century.

This paper reviews the development of high-power microwave sources based on vacuum tube technology. The following chapters are classified according to the principle of vacuum tube microwave generation, and various types of vacuum tubes are discussed separately. In fact, so far, there are more than 2,000 papers related to gyrotrons alone [18]. It is impossible for us to describe all related studies in detail, and only discuss the most important basic principles and progress.

## 2. Cherenkov or Smith-Purcell Radiation Devices

Cherenkov radiation is a kind of short-wavelength electromagnetic radiation emitted by charged particles when it was moving in a medium and its velocity was faster than the speed of light in the medium, which was discovered by former Soviet physicist Cherenkov in 1934 [19]. Smith-Purcell radiation is the phenomenon of radiated light when free electrons sweep across the grating surface discovered by S. J. Smith and E. M. Purcell in 1953 [20]. These two radiation principles are similar [21] and can be attributed to the role of slow-wave structures.

In a periodic system, electrons interact with electromagnetic waves [22], when the direction of electromagnetic wave propagation is the same as the direction of electron movement,

$$\omega - k_z v_z \simeq \mathrm{n}\bar{k} v_z, \tag{1}$$

where, $\omega$ is the angular frequency, $k_z$ is the longitudinal wave number, $\bar{k} = 2\pi/L$, $L$ is the period length of the slow wave structure. When the electron beam interacts with the fundamental wave, n=0; When the electron beam interacts with higher harmonics, n≠0.

When the direction of electromagnetic wave propagation is opposite to the direction of electron movement (counterpropagating wave),

$$\omega + k_z v_z \simeq -\mathrm{n}\bar{k} v_z. \tag{2}$$

The most common Cherenkov devices are TWT and BWO. The Brillouin diagram is shown in Fig. 2. The difference between the maximum frequency at the $\pi$ point and the minimum frequency at the 0 point depends on the depth of the slow wave structure ripple [23].

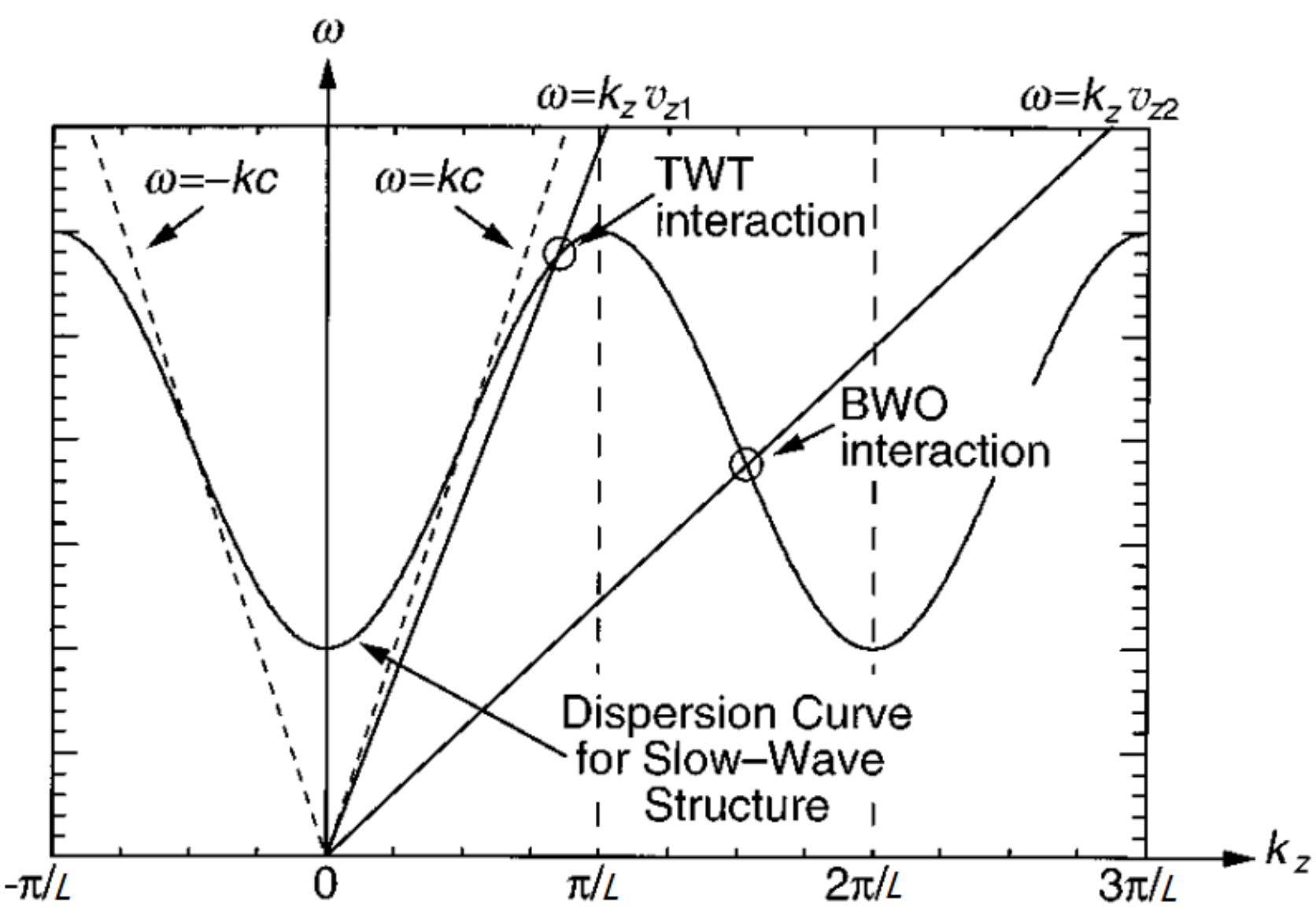

FIG. 2. The Brillouin diagram of the TWT and the BWO.

Orthogonal field devices such as magnetrons and MILOs are different from linear devices such as TWTs. When electrons drift in a resonant structure, they convert the potential energy of the electrons into microwaves. However, since the drift speed of the electron is close to the phase velocity of the slow wave, they can still be regarded as the Cherenkov devices [10].

## 2.1. Traveling Wave Tube (TWT)

Most literature believes that the traveling wave tube was invented in the United Kingdom in 1943 by Australian-born engineer Rudolf Kompfner. In fact, Andrei Haeff of the United States also made important contributions. It can be said that he invented the prototype of the first traveling wave tube and applied for a patent in 1933. Related information can be found in the IEEE Spectrum article 'Andrei Haeff and the Amazing Microwave Amplifier' [24].

After decades of development, some microwave electric vacuum device R&D companies in developed countries in Europe and the United States already have many mature products, which are widely used in missile guidance, electronic countermeasures, radar, telecommunications [25], and other fields.

The traveling wave tube is mainly composed of an electron gun, a slow wave structure, a magnet (a wire package or a periodic permanent magnet structure), and a collector. The electron gun emits an electron beam, and the microwave is amplified by the interaction of the slow wave structure, and the collector is used to absorb the remaining electron beam. The slow-wave structures are in the form of spirals, coupling cavities, folded waveguides, etc. [26].

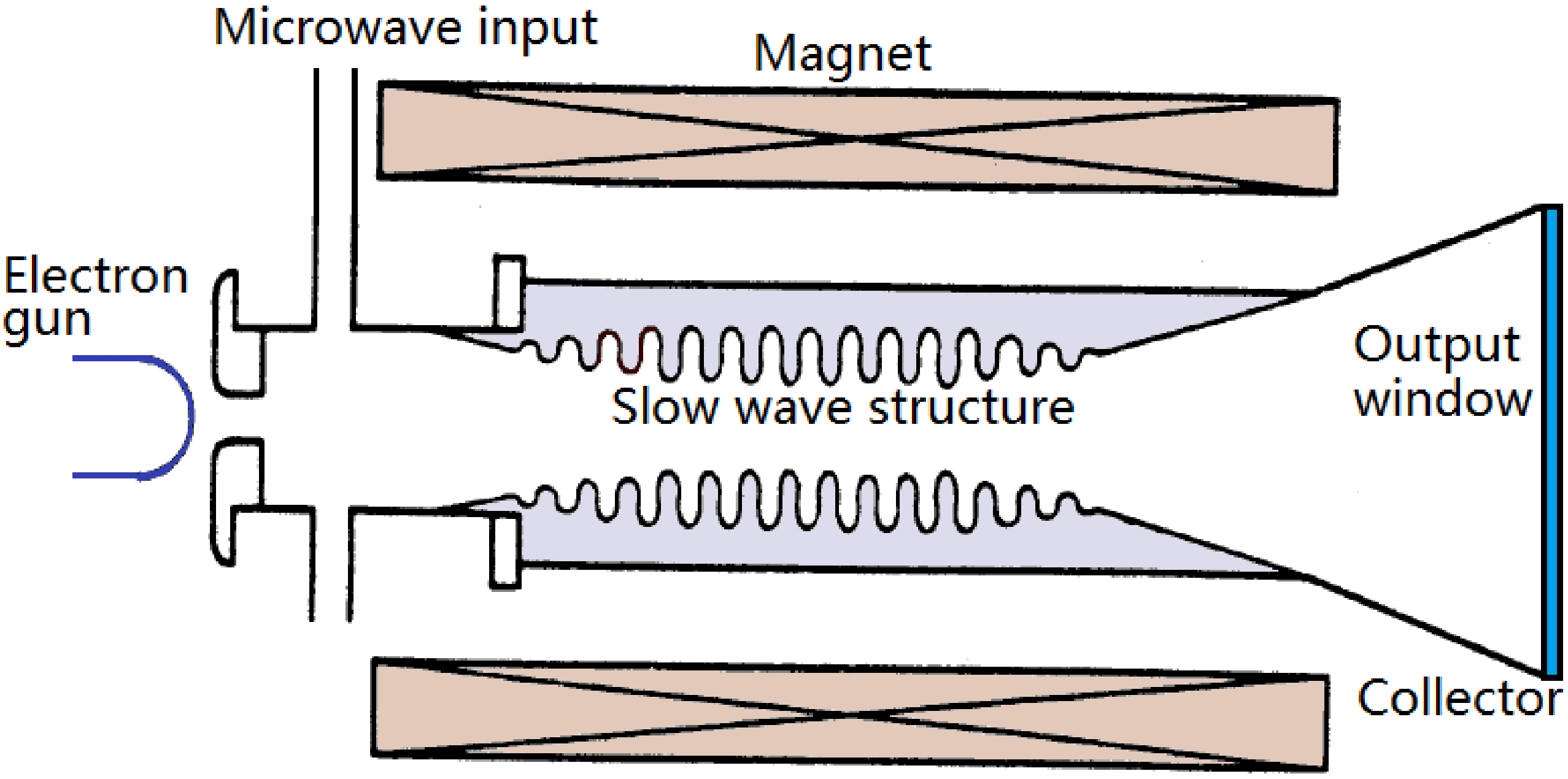

FIG. 3. The basic structure of the traveling wave tube.

Designing and analyzing the electron-optical system is of primary importance in designing a traveling wave tube. Using finite element, finite difference and other computational electromagnetic technology and PIC technology, a large amount of electron-optical CAD software has been developed, and the interaction between electromagnetic waves and electrons can be simulated and analyzed using a computer. In addition, the lumped circuit model can also be used to analyze the traveling wave tube. The more widely used is the Pierce equivalent circuit model. In order to solve the problem that the Pierce impedance tends to infinity at the cutoff frequency,

Damien F G Minenna et al. established a discrete model that can be used for small signal analysis of traveling wave tubes [27].

The working frequency of the traveling wave tube can be from L band to THz band. Some typical research progress is shown in Table II and III. The development of solid-state devices has a huge impact on vacuum devices such as traveling wave tubes in the centimeter wave band and lower frequency and low power fields. Therefore, at present, researches on traveling wave tubes are mainly focused on high frequency (millimeter wave, THz), high power (tens, hundreds of kilowatts and above), and high bandwidth.

Millimeter wave and THz wave traveling wave tubes have been extensively studied. Among them, the folded waveguide slow-wave structure is more widely used because it is easier to fabricate than the spiral and coupled cavity, the all-metal structure is resistant to high power consumption, and the bandwidth is relatively wide [28]. However, the traveling wave tube of this structure has low coupling resistance and large attenuation, and its efficiency is generally low. John H. Booske et al. established a parametric model [29] that can be used to simulate folded waveguide millimeter wave traveling wave tubes. At present, the research focus of millimeter wave traveling wave tubes includes slow wave structures [30], electron-optical systems, etc., in order to improve power and bandwidth. In order to increase the flexibility of traveling wave tubes, traveling wave tubes with adjustable power and frequency have also been studied.

TABLE II. Research progress of L ~ V band traveling wave tube.

| **Band** | **Peak Power [W]** | **Freq. [GHz]** | **Gain [dB]** | **Efficiency [%]** | **Pulse length [s]** | **Duty cycle [%]** | **Institute** | **Model** |
|---|---|---|---|---|---|---|---|---|
| L | 170k | 1.2-1.4 | | 28 | 100µ | 5 | TMD | PT6049 |
| L | 12k | 1.75-1.85 | | | CW | | CPI | VTL-6640 |
| S | 170k | 3.1-3.5 | | 22 | 1000µ | 16 | CPI | VTS-5753 |
| S~C | 150[25] | 3.4–4.2 | 50 | 73 | CW | | TED | TL4150 |
| C | 40k | 5.85-6.425 | | 11 | CW | | CPI | VTC-6660E |
| X | 100k | 8. 5-9. 6 | | 15 | 100µ | 3.3 | TED | TH3897 |
| X | 15k | 10.1-10.7 | | | CW | | CPI | VTX-6383 |
| X~Ku | 150 | 10.7–12.75 | > 50 | 68 | CW | | TED | TH4795 |
| Ku | 60k | 15.7-17.7 | | 11 | 300µ | 30 | CPI | VTU-5692C |
| Ku~K | 160 | 17.3–20.2 | > 50 | 63 | CW | | TED | TH4816 |
| K | 50 | ~26 | > 50 | 55 | CW | | TED | TH4626 |
| Ka | 500 | 28.3-30 | 37-48 | 29 | CW | | CPI | VTA-6430A2 |
| Ka | 560 | 30 | 42.5 | 16 | CW | | TED | LD7319 |
| Ka | 35 | 32 | > 50 | 54 | CW | | TED | TH4606C |
| Ka | 50k | 34.5-35.5 | | | | 10 | CPI | VTA-5710 |
| Ka~V | 40 | 37.5–42.5 | 48 | 50 | CW | | TED | THL40040CC |
| V | 100 | 43.5-45.5 | | 30 | CW | | TED | TH4034C |
| V | 230 | 43.5-45.5 | 41 | | CW | | L3 | 8925HP |
| V | 21[31] | 55-60 | 38 | 8.5 | | | TTEG | |

Note: TMD [32] is the abbreviation of TMD Technologies LLC in the United States. CPI is the abbreviation of Communications & Power Industries LLC in the United States. TED is the abbreviation of Thales Electronic Device in France. L3 is the abbreviation of Communications Electron Technologies Incorporation (L3-ETI) in the

United States. TTEG is the abbreviation of Thomson Tubes Electroniques GmbH, Germany.

TABLE III. Research progress of millimeter wave and THz band traveling wave tubes.

| Band | Average power [W] | Peak power [W] | Center Freq. [Hz] | 3dB band width [Hz] | Gain [dB] | Efficiency [%] | Pulse length [s] | Duty Cycle [%] | Institute | Model |
|---|---|---|---|---|---|---|---|---|---|---|
| W | 50 | | 93~95G | | | | CW | | CPI | VTW6495 |
| W | | 3k[33] | 95~96G | | | | | 10 | CPI | VTW-5795 |
| W | | 3k | 93.9~94.1G | | | | | 10 | CPI | VTW-5795A2 |
| W | | 200 | | | | | | 10 | TTEG | |
| W | | 150 | 93~95G | | | 15 | | 10 | TED | TH4402-1 |
| W | | 100 | 93~95G | | | 15 | | 20 | TED | TH4402-2 |
| W | 100 | 300 | 90.6G | | 30 | | | | L3 | |
| W | 100 | 200 | 91.4G | | 30 | | | | L3 | |
| W | | >60[34] | 94~110G | | >30 | >1.7 | | 1 | CETC 12th institute | |
| W | | >250[28] | 89.6~97.6G | | >30 | >5.8 | | 1 | CETC 12th institute | |
| mm | 35.5 | 79[35] | 233G | 2.4G~3G | 23 | 2 | | 50 | NGC | |
| THz | | 259m[36] | 640G | 15G | 22 | | | 10 | NGC | |
| THz | | 71m | 670G | 15G | 17 | | | 0.5 | NGC | |
| THz | | 39m | 850G | 15G | 22 | | | 11 | NGC | |
| THz | | 29m | 1030G | 5G | 20 | | | 0.3 | NGC | |

Note: NGC is the abbreviation of Northrop Grumman Corporation in the United States. CETC 12th institute is the abbreviation of China Electronics Technology Group Corporation 12th Research Institute, i.e. Beijing Vacuum Electronics Research Institute.

In the 1980s, relativistic TWTs using relativistic electron beams began to develop. In 1990, Donald Shiffler and others of Cornell University reported an X-band corrugated waveguide relativistic TWTs. It uses a field emission cathode as a pulse energy source, and an electron beam is generated by a field emission cathode immersed in a magnetic field. The electron beam has a voltage of up to 850 kV and a current of 1 kA. It has gained a gain of 13 ~ 35 dB at 8.76 GHz, and its output power range is 3 ~ 100 MW. The amplifier is designed to operate in the narrow band of $TM_{01}$ mode [37]. China's Northwest Institute of Nuclear Technology (NINT) and other research institutes have also conducted research on relativistic TWTs. Table IV lists some research progress of relativistic TWTs. The study found that plasma filling in microwave devices can improve performance and increase output power. The University of Maryland, the University of Electronic Science and Technology of China, etc. have conducted theoretical and experimental research on plasma-filled traveling wave tubes and made some progress [38,39].

TABLE IV. Research progress of relativistic TWTs.

| Band | Peak power [W] | Freq. [GHz] | Gain [dB] | Efficiency [%] | Pulse length [s] | Duty cycle [%] | Beam voltage [V] | Beam current [A] | Institute |
|---|---|---|---|---|---|---|---|---|---|
| X | 3M-100M | 8.76 | 13-35 | 11 | 100n | | 850k | 1k | Cornell University |
| X | 70M[40] | 9 | | 20 | 100n | | 700k | 500 | Cornell University |
| X | 2.1G[41] | 9.3 | | 41 | 10n | | 680k | 7.5k | NINT |
| X | 1.2G | 10.3 | | 35 | 10n | | 580k | 5.9k | NINT |
| mm | 70k (simulation result)[42] | 220 | 28.5 | | | | | | CETC 12th institute |

## 2.2. Backward Wave Oscillator (BWO)

A backward wave oscillator is also a Cherenkov device. Unlike a traveling wave tube, in a backward wave oscillator, the wave group velocity is opposite to the direction of electron movement, so it is called a backward wave oscillator (or a backward wave tube). The BWO is the first experimentally verified high-current relativistic high-power microwave source. In 1973, it produced 400 MW, 10 ns high-power microwave output in the former Soviet Union [43].

The BWO structure is similar to that of the TWT, as shown in Fig. 4 and is mainly composed of an electron gun, a magnet, a slow-wave structure, a Bragg reflector, a collector, and an output window. The Bragg reflector is used to reflect microwaves to the output port, and can also be replaced by cut-off waveguides, resonant reflectors, etc. There is also a type of BWO with microwave output near the end of the electron gun. A BWO with this structure sometimes installs a terminal absorber near the terminal of the slow-wave structure of the collector to absorb the forward wave and prevent the wave reflection from adversely affecting the tube.

J. A. Swegle et al. established the linear theory of the BWO [44-47], and gave a basic analysis of the beam-wave interaction of the BWO. B. Levush established a set of nonlinear equations describing the beam-wave interaction of the BWO, and considered the effects of reflections at both ends [48]. In the slow-wave structure of the BWO, the electron beam generates velocity modulation and density modulation under the action of microwaves, and the clustered electrons transfer energy to the microwave in the deceleration region of the high-frequency field. In contrast to the group velocity, the phase velocity of the wave and the direction of the electron beam are the same in the BWO, and their velocity is close to equal. Due to the dispersion of the return wave in the slow-wave structure, when the electron beam voltage is changed (that is, the electron beam velocity is changed), the oscillation frequency that satisfies the phase condition changes accordingly. In general, due to the dispersion characteristics of its slow-wave structure with $dv_p/d\omega>0$ (where $v_p$ is the phase velocity), the oscillation frequency increases with the increase of the electron beam voltage. When a stable frequency is required, an external reference can be used to achieve phase lock [49,50].

At present, there are two main research directions of BWOs. One is the study of high-frequency (millimeter wave and THz wave) BWOs, and the other is the research of high-power relativistic BWOs.

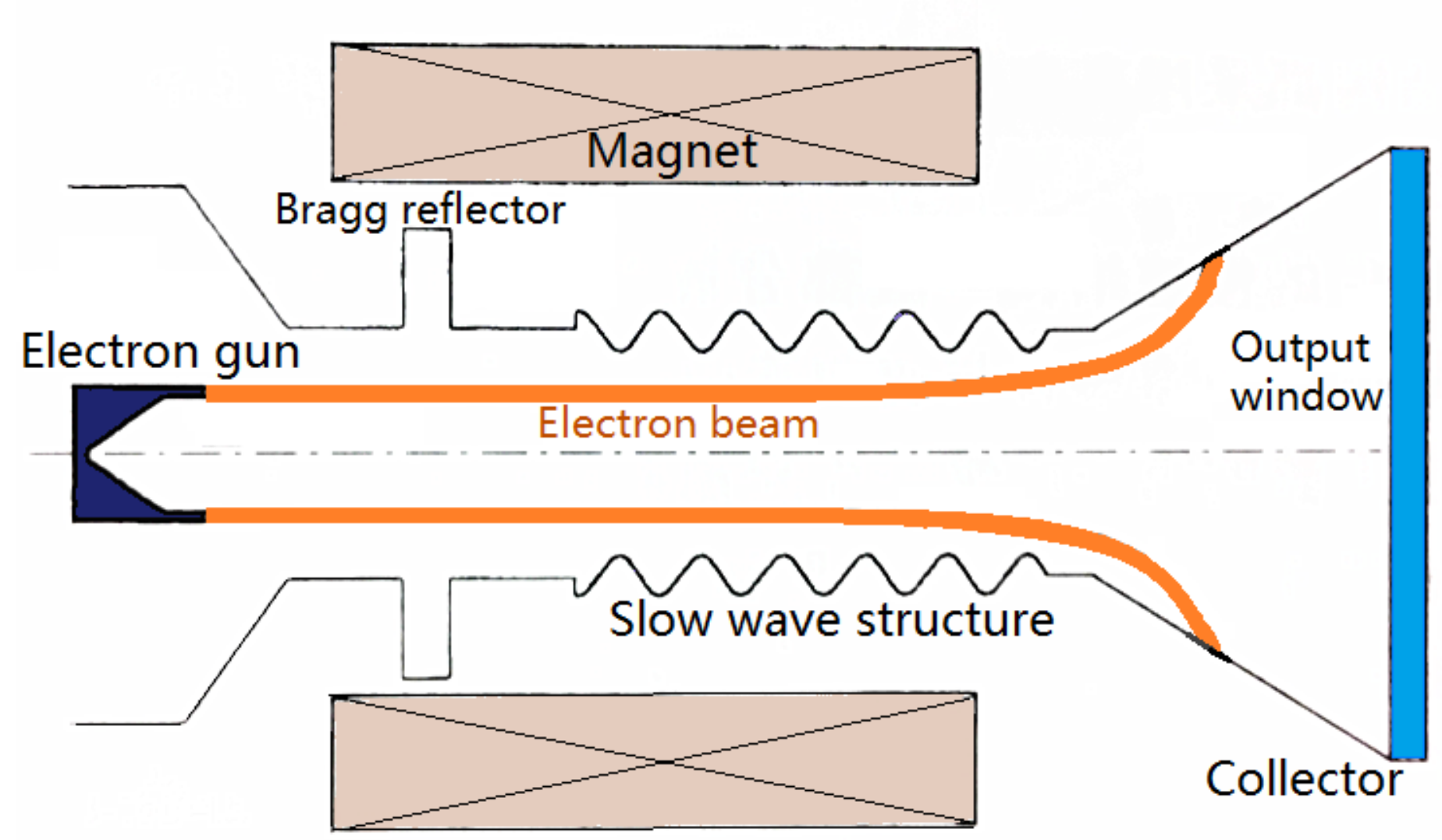

FIG. 4. The basic structure of the BWO.

The research institutes for millimeter wave and terahertz BWOs include Russia's ISTOK company, the US CCR company, the US MICROTECH company, and the CETC 12th institute. Among them, the ISTOK has developed the BWO with a frequency covering 36 GHz to 1400 GHz [51]. Table V lists the research progress of non-relativistic millimeter wave and terahertz BWOs by research institutions represented by ISTOK.

Similar to the millimeter wave TWT, in the millimeter wave BWO, a folded waveguide is mostly used as a slow wave structure. The main research directions of millimeter wave and THz BWOs include increasing frequency, using multi-stage depressed collectors and other technologies to improve efficiency, improve output coupling efficiency, reduce the required magnetic field or reduce the volume of magnet system . Progress has also been made in the research of the multi-frequency BWOs [52] and the plasma-filled BWOs [39].

TABLE V. Research progress of non-relativistic millimeter wave and terahertz BWOs.

| **Band** | **Power [mW]** | **Freq. [GHz]** | **Efficiency [%]** | **Pulse length [s]** | **Beam voltage [V]** | **Beam current [mA]** | **Institute** | **Model** |
|---|---|---|---|---|---|---|---|---|
| Ka~V | 15-40 | 36-55 | ~0.1 | | 400-1200 | 20-25 | ISTOK | OB-69 |
| V~W | 12-30 | 52-79 | ~0.1 | | 400-1200 | 20-25 | ISTOK | OB-70 |
| W~mm | 6-30 | 78-119 | ~0.05 | | 500-1500 | 20-25 | ISTOK | OB-71 |
| mm | 6-20 | 118-178 | ~0.05 | | 500-1500 | 20-25 | ISTOK | OB-86 |
| mm | 6-15 | 177-260 | ~0.05 | | 700-1900 | 15-22 | ISTOK | OB-24 |
| mm~THz | 1-10 | 258-375 | ~0.005 | | 1000-4000 | 25-40 | ISTOK | OB-30 |
| THz | 1-5 | 370-535 | ~0.005 | | 1000-4500 | 25-40 | ISTOK | OB-32 |
| THz | 1-5 | 530-714 | ~0.002 | | 1500-6000 | 30-45 | ISTOK | OB-80 |
| THz | 1-5 | 690-850 | ~0.002 | | 1500-6000 | 30-45 | ISTOK | OB-81 |
| THz | 0.5-5 | 790-970 | ~0.001 | | 1500-6000 | 30-45 | ISTOK | OB-82 |
| THz | 0.5-3 | 900-1100 | ~0.001 | | 1500-6000 | 30-45 | ISTOK | OB-83 |
| THz | 0.5-2 | 1070-1200 | ~0.001 | | 1500-6000 | 30-45 | ISTOK | OB-84 |
| THz | 0.5-2 | 1170-1400 | ~0.001 | | 1500-6000 | 30-45 | ISTOK | OB-85 |
| THz | 10 (simulation | 337-347 | ~0.01 | | 14000-16200 | 8 | CETC | |

| | | | | | | | |
|---|---|---|---|---|---|---|---|
| | result) [53] | | | | | | $12^{th}$ institute |
| THz | >8k (simulation result) [54] | 340-368.2 | ~3 | CW | 11k-17k | 20 | IECAS |

Note: IECAS is short for Institute of Electronics, Chinese Academy of Sciences.

The relativistic BWO has the characteristics of high power, high efficiency, and simple structure. The relativistic BWOs are widely used in high-power microwave weapons and other fields. The institutions that study relativistic BWOs include the Russian Institute of Applied Physics (IAP), the Russian Institute of High Current Electronics (IHCE), the University of Maryland, Cornell University, the University of New Mexico, the China's Northwestern Institute of Nuclear Technology, the China's National University of Defense Technology (NUDT), and the China Academy of Engineering Physics (CAEP).

In order to improve power, a large over-mode slow-wave structure is generally used. So far, the maximum peak power has exceeded 5 GW in S-band, X-band and other frequencies. For a small over-mode BWO ($D/\lambda$ is about 1.8, D is the diameter of the interaction zone, $\lambda$ is the free-space wavelength), at 550kV, the maximum power obtained is 0.8 GW and the frequency is 10 GHz [55]. The research progress of the relativistic BWOs is shown in Table VI. It is worth noting that the efficiency of some ultra-short pulse relativistic BWOs exceeds 100%. This is the use of the spatial accumulation effect of energy in ultra-short microwave pulses, which produces pulses with peak power significantly higher than the power of the electron beam. Related progress can refer to literature [56].

The main disadvantage of the relativistic back wave oscillator is that it generally requires a large magnetic field (> 2T), and the magnet system is too large. Therefore, one of the research directions now is the relativistic BWOs with the low guiding magnetic field. In addition, higher power, higher pulse width, and higher efficiency BWOs are also important research directions. To increase the pulse width, the pulse shortening effect needs to be overcome [57]. For example, Xingjun Ge of the National University of Defense Technology used two cavities to reduce the RF field, and introduced a large-radius collector to reduce the number of secondary electrons generated by electron bombardment. The microwave output power was 2 GW with 110 ns pulse length in the S-band [58].

TABLE VI. The research progress of the relativistic BWOs.

| **Band** | **Peak power [W]** | **Freq. [Hz]** | **Efficiency [%]** | **Pulse length [s]** | **Beam voltage [V]** | **Beam current [A]** | **Institute** |
|---|---|---|---|---|---|---|---|
| L | 1.05G[59] | 1.61G | 14.4 | 38n | 703k | 10.6k | NUDT |
| S | 1G[60] | 3.6G | 20 | 100ns (can run at 10Hz repeat frequency) | 700k | 7k | NUDT |
| S | 5.3G[61] | 3.6G | 30 | 25ns | 1.2M | 15k | IHCE |
| S | 2G[58] | 3.755G | 30 | 110n | 820k | 8.1k | NUDT |
| X | 5.06G[62] | 8.25G | 25 | 13.8n | 1M | 20k | CAEP, Tsinghua University |
| X | 0.8G[55] | 10G | 24 | | 550k | 6k | IHCE |
| X | 0.55G[63] | 9.45G | 17 | 8n | 620k | 5.2k | The University of New Mexico |
| X | 0.9G[64] | 9.4G | 29 | 32ns | 500k | 6.2k | SWUST, CAEP |
| X | 1.4G[65] | 9.4G | 26 | 30ns | 790k | 6.7k | CAEP |
| X | 3G[66] | | 7.5 | 30n | 2M | 20k | IHCE |
| Ka | 1.1G[67] | 38G | ~150 | 0.2n | 290k | 2.3k | IERAS, IHCE |
| Ka | 400M[68] | 38G | 66 | 0.2n | 290k | 2.1k | IERAS, IHCE |
| mm | 100k[69] | 140G | 0.1 | 1-2n | 100k | 1k | CAEP |
| THz | 600k[70] | 340G | 0.6 | 1.66n | 100k | 900 | CAEP |

Note: SWUST is short for China's Southwest University of Science and Technology. IERAS is short for Institute of Electrophysics, Russian Academy of Sciences.

## 2.3. Surface Wave Oscillator

Surface wave oscillator (Cerenkov Oscillator With the Bragg Reflection Resonator) is a very powerful high-power microwave and terahertz wave source. In order to improve the power capacity, a large-size over-mode waveguide is used [71]. The maximum value of the electromagnetic wave electric field appears near the surface of the slow wave structure, so it is called the surface wave oscillator. For a smooth-walled cylindrical waveguide, for $TM_{mn}$ mode, the relationship between the allowable maximum power $P_{max}$ and the maximum electric field strength $E_{max,w}$ near the waveguide wall is [72],

$$P_{max}[\mathrm{GW}] = 8.707\left(1+\delta_{0,m}\right)\left(\frac{E_{max,w}\left[\frac{\mathrm{kV}}{\mathrm{cm}}\right]\lambda[\mathrm{cm}]}{511}\right)^2 \times \frac{\pi^2}{4} D'^4 \sqrt{1-\left(\frac{v_{m,n}}{2\pi D'}\right)^2}\,\frac{1}{v_{m,n}^2}. \qquad (3)$$

Where, $D' = \pi D/\lambda$, D is the inner diameter of the waveguide, $\lambda$ is the free space wavelength, $\nu_{m,n}$ is the root of the m-order Bessel Function $J_m(x)=0$. When m=0, $\delta_{0,m}=1$, otherwise $\delta_{0,m}=0$. Using an over-mode waveguide can increase D', thereby increasing power capacity.

The surface wave oscillator generally operates in the fundamental mode $TM_{01}$. The intersection of the Doppler line of the electron beam $\omega=k_z v_z$ and the dispersion curve of the $TM_{01}$ mode is located to the left near the point $\pi$, as shown by the intersection of the $V_{p2}$ Doppler line and $TM_{01}$ mode dispersion curve shown in Fig. 5. The group velocity of the generated wave is positive, and the phase velocity is also positive, which is in a traveling wave state [73].

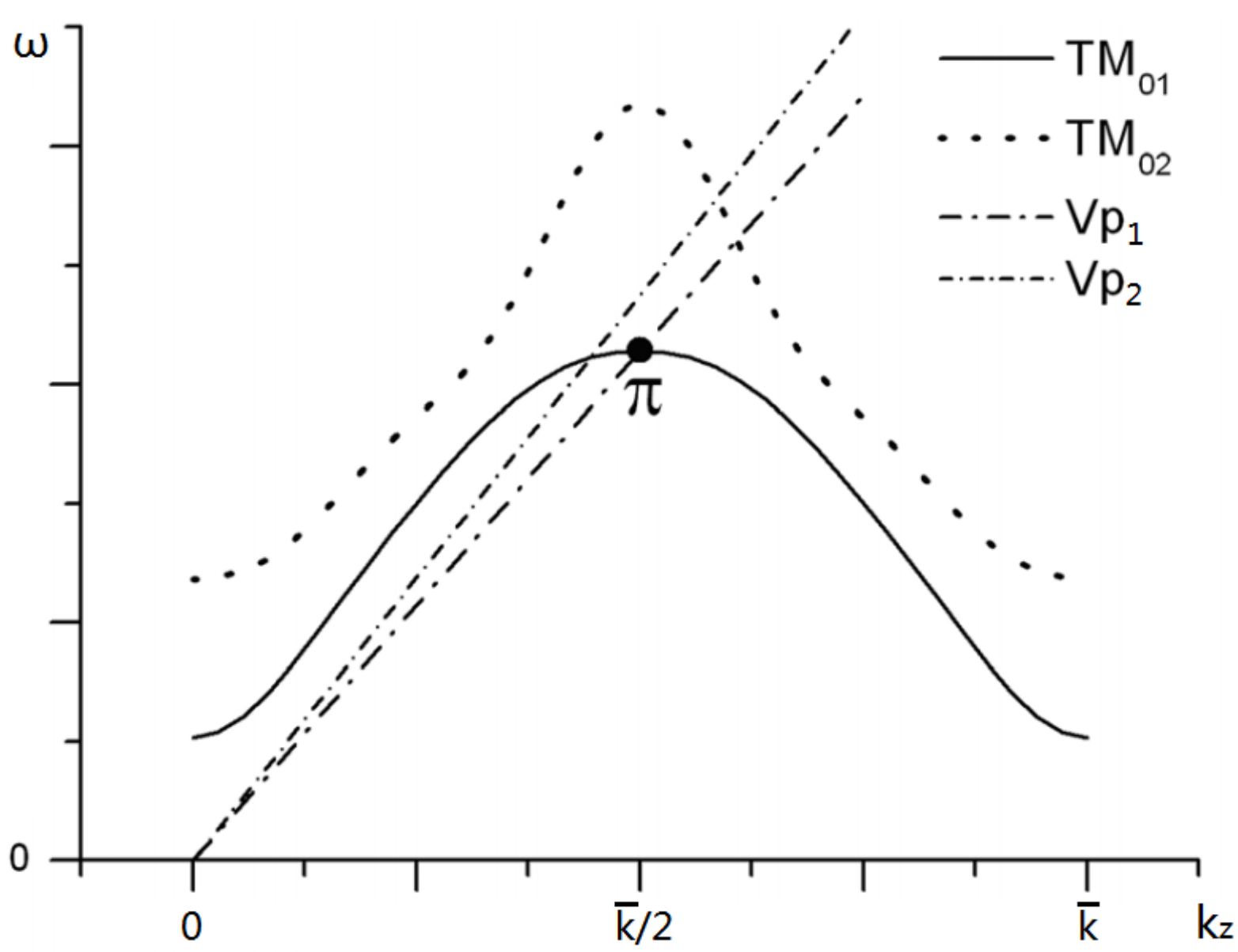

Fig. 5. Brillouin plot of a surface wave oscillator, where $k_z$ is the longitudinal wave number.

In the millimeter wave band, many simulations and experiments have been done by China's Northwest Institute of Nuclear Technology, Russian Institute of Applied Physics, etc. Most of the simulation results are good, but the experimental results are very inefficient. Moreover, in the millimeter wave band, the output mode of the surface wave oscillator is not pure, and there are many high-order modes mixed [22,74].

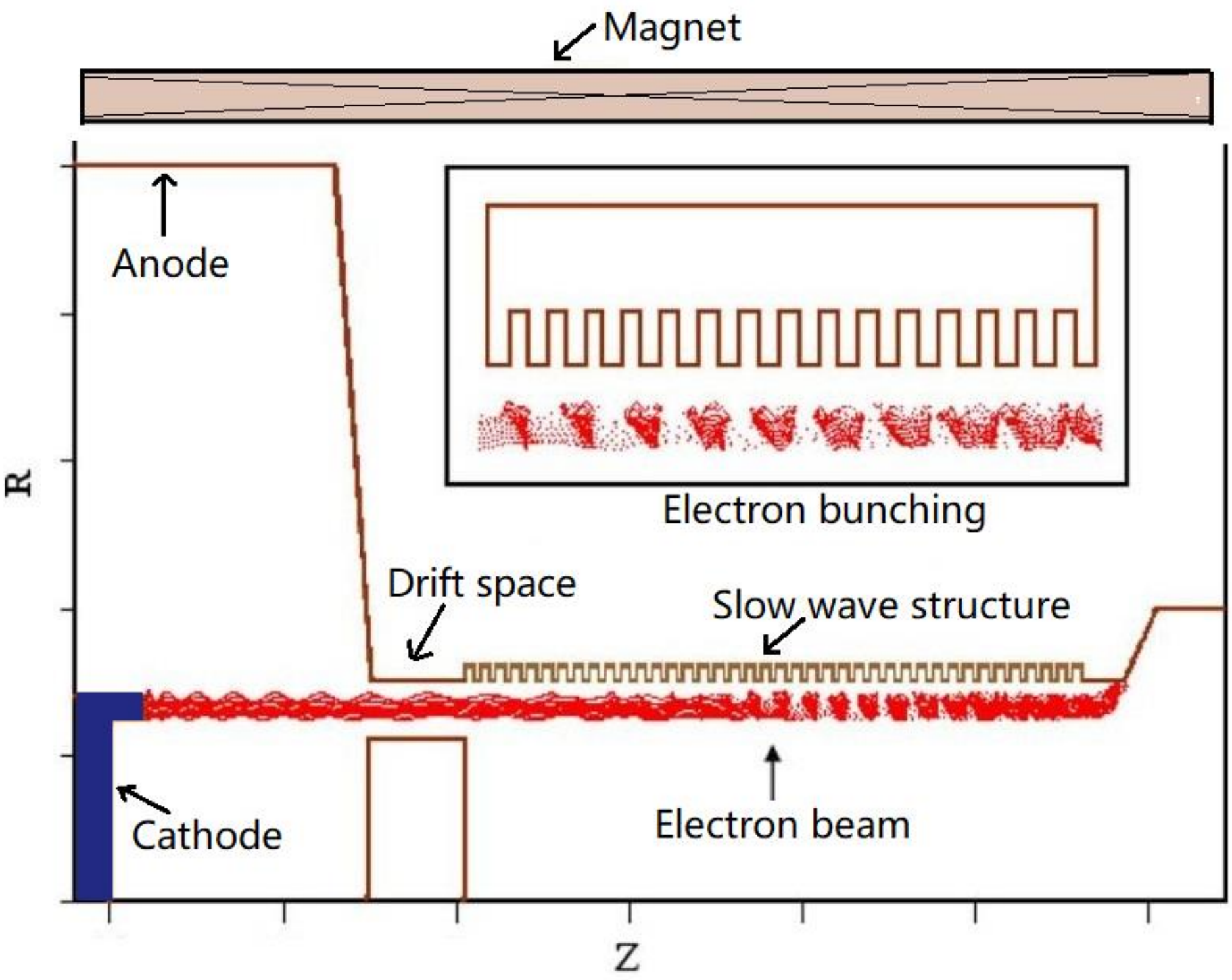

Fig. 6. The basic structure of the surface wave oscillator and the schematic diagram of electron clustering.

TABLE VII. Advances in simulation and experimental research of surface wave oscillators.

| Band | Peak power [MW] | Freq. [GHz] | 1 dB bandwidth | Efficiency [%] | Pulse length [ns] | Institute | Years | Research form |
|---|---|---|---|---|---|---|---|---|
| X | 500 [72] | 8.3 | | 15 | 20 | University of Maryland | 2000 | Experiment |
| Ka | 50 [22] | 33.3 | | 6 | 15~30 | IAP | 1984 | Experiment |
| E | 12 | 62.5 | | 4 | 10~15 | IAP | 1984 | Experiment |
| D | 8 | 125 | | 3 | 5~10 | IAP | 1984 | Experiment |
| D | 680 [75] | 140 | | 90.67 | 0.11 | Xi'an Jiaotong University | 2009 | Simulation |
| D | 2.6 [74] | 154 | | 1.4 | 1.5 | NINT | 2013 | Experiment |
| D | 5 [71] | 148 | | 0.7 | 3 | NINT | 2013 | Experiment |
| mm | 0.5 [22] | 333 | | 0.2 | 3~5 | IAP | 1984 | Experiment |

## 2.4. Multi-Wave Cherenkov Generator (MWCG)

Multi-wave Cherenkov generator (MWCG for short) is one of the most powerful microwave devices. Similar to surface wave oscillator, in order to obtain high output power, an over-mode structure is used. MWCG uses a two-stage slow wave structure with the same space period and requires a strong axial magnetic field. Both sections of the slow-wave structure work near the $\pi$ point of the dispersion curve. Its basic structure is shown in Fig. 7 [76].

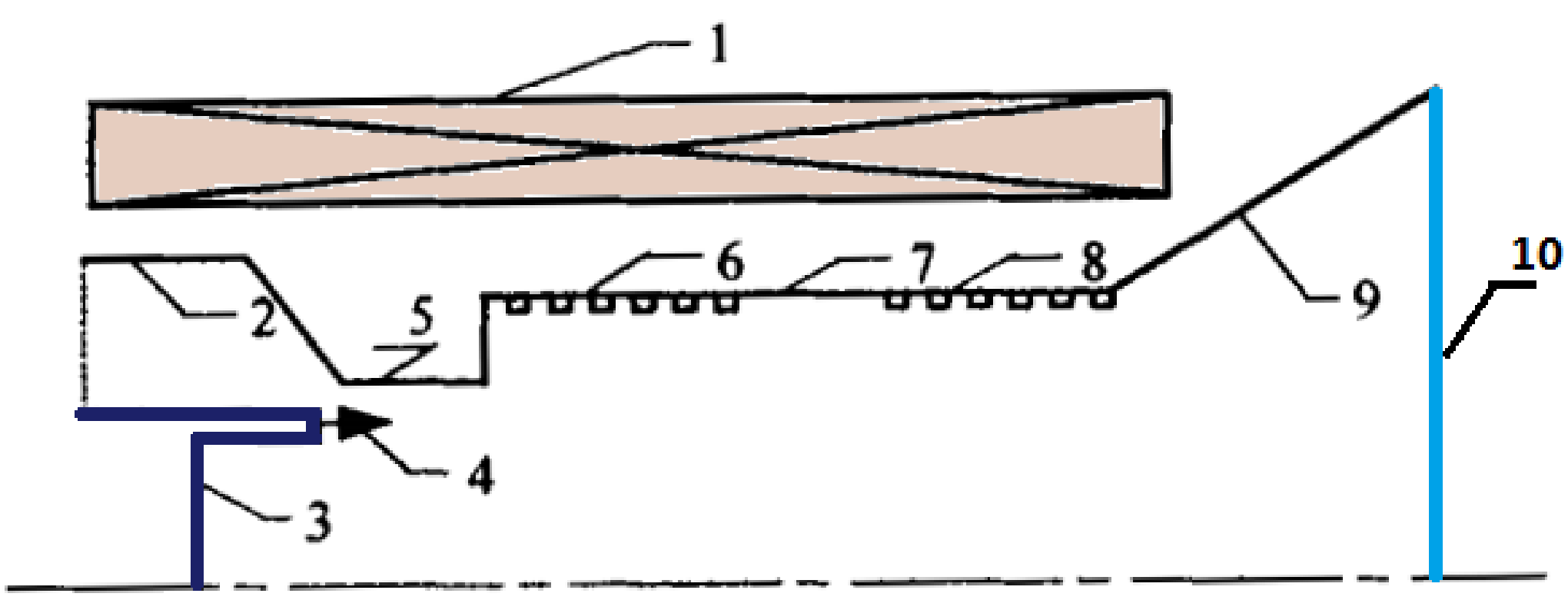

FIG. 7. The basic structure of MWCG. 1 is the magnetic field coil, 2 is the anode, 3 is the cathode, 4 is the relativistic electron beam, 5 is the beam collimator, 6 is the first-stage slow-wave region, 7 is the fast-wave drift tube section, 8 is the second-stage slow wave region, 9 is a horn structure that absorbs residual electron beams, and 10 is a dielectric output window [77].

The electron beam undergoes speed modulation in the first-stage slow wave region; then it is converted into density modulation in the drift tube section (fast wave region) to form an electron pre-grouping; In the second-stage slow-wave region, the electron beam interacts with electromagnetic waves efficiently, so the high-power microwaves are generated and output through the output window. Usually, the two-stage slow-wave structures are symmetrical, and there are also studies of MWCG using asymmetric slow-wave structure [78].

The horizontal and vertical dimensions of the slow-wave structures of each segment of MWCG are close to each other, and surface waves cannot be generated like the SWOs. The electron beams interact with surface waves and body waves simultaneously inside the MWCGs. Due to the open electromagnetic structure and complex electron beam physical process of MWCGs, there is no complete self-consistent non-linear theory to analyze the interaction process. The approximate linear theory can be used to analyze the electromagnetic field, the starting current, etc. [79].

A large number of theoretical and experimental studies on MWCG have been conducted by research institutions represented by the Russian Institute of High Current Electronics since the 1980s [80]. Some research progress is shown in Table VIII. At present, the MWCG can generate a maximum of 15GW in the 3cm band. In recent years, combined with the advantages of other microwave tubes such as klystrons, research on new devices such as Cherenkov generators in the form of klystrons has appeared [81].

TABLE VIII. The progress in experimental research of MWCGs.

| Band | Peak power [GW] | Freq. [GHz] | Efficiency [%] | Pulse length [s] | Institute | Years |
|---|---|---|---|---|---|---|
| X | 0.2 | 8.8-9.77 | 10 | 0.5μ-0.6μ | IHCE | 1990 |
| X | 15 | 10 | 50 | 60n~70n | IHCE | 1990 |

| X | 0.85 [78] | 9.23 | 30 | 30n | IHCE | 2013 |
|---|---|---|---|---|---|---|
| Ka | 3 | 30.86 | 20 | 60n~80n | IHCE | 1990 |
| Ka | 1.5 [79] | 34.80 | 15 | 60n~80n | IHCE | 1990 |
| Ka | 0.5~0.6 [82] | 33.94 | 6~7 | 20n | IHCE | 2000 |

## 2.5. Relativistic Diffraction Generator (RDG)

The relativistic diffraction generator (RDG) also uses an over-mode structure. In an RDG, the microwave field is distributed in structural space, as opposed to devices based on interactions with surface waves (SWO, MWCG). The working range of the RDGs falls within the frequency region of 2π-type oscillations of the lower axially symmetric mode of the periodic waveguide [83].

At present, the highest peak power is recorded at 9 GW, and the wavelength is 9-11.3 mm [84].

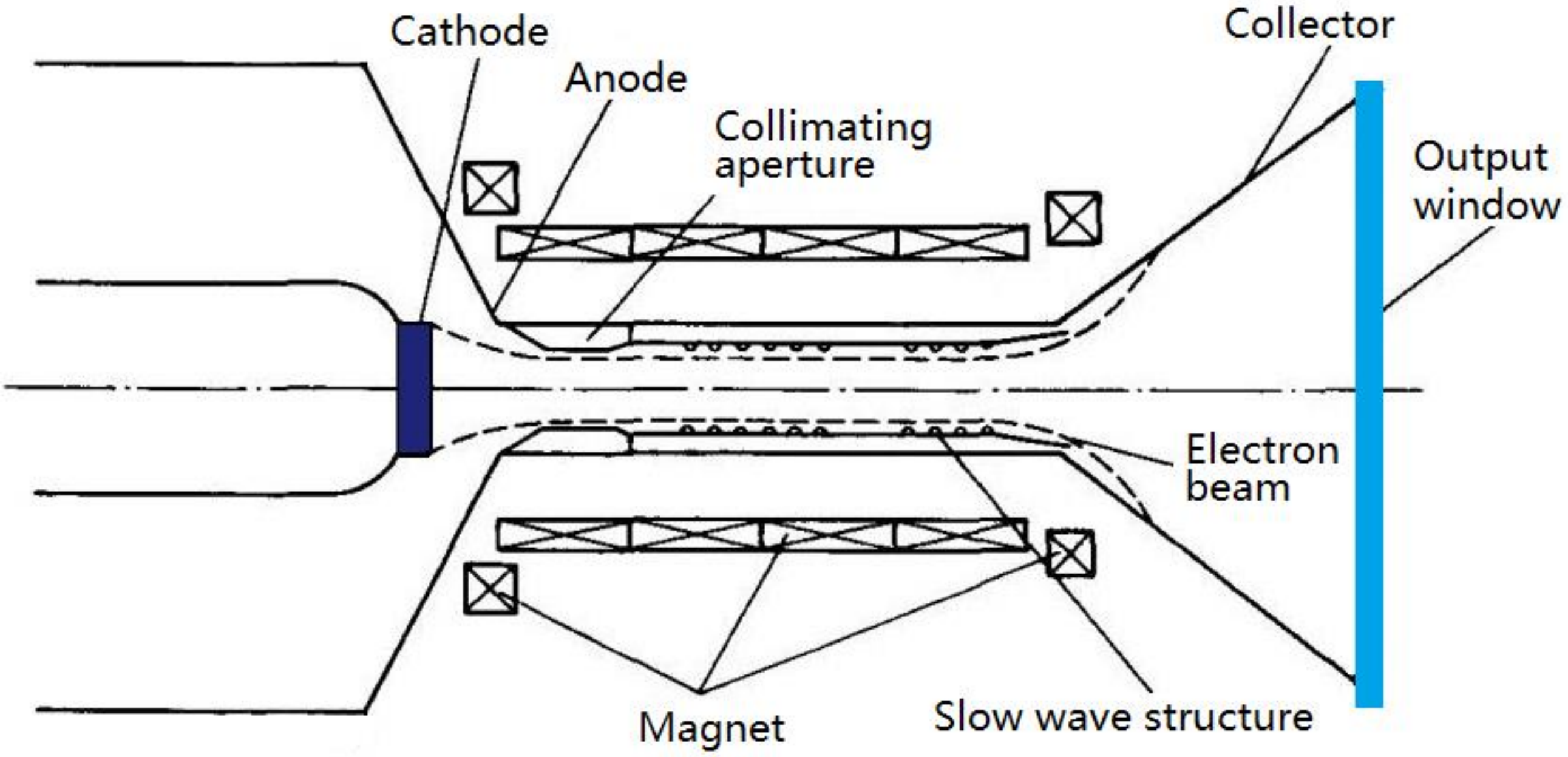

FIG. 8. The basic structure of RDG.

TABLE IX. The progress in experimental research of RDGs.

| Band | Peak power [GW] | Freq. [GHz] | Efficiency [%] | Pulse length [s] | Beam voltage [V] | Beam current [A] | Institute | Years |
|---|---|---|---|---|---|---|---|---|
| K-Ka | 9 [84] | 26.55-33.33 | 33 | 100n-350n | 1.6M | 17k | IHCE | 1990 |
| V | 2 | 41.7 | | | | | IHCE | 1990 |
| V | 7 | 44.1-46.2 | 29 | 200n-260n | 1.5M | 16k | IHCE | 1990 |
| V | 5.6 | 46.2 | 17 | 700n | 1.7M | 19k | IHCE | 1990 |

## 2.6. Orotron

Orotron is also called 'diffraction radiation oscillator' or 'laddertron' [85]. The horizontal field of the Orotron is much larger than the vertical field ($k_{\perp} \gg k_z$), and it works in the first harmonic state.

$$\omega \simeq \bar{k} v_z. \qquad (4)$$

Relative to the π mode of an SWO, the Orotron oscillation is called 2π mode. A linear theory was proposed by Richard P. Leavitt et al [86] to calculate the starting current and the electronic tuning characteristics. A nonlinear theory showing how to optimize the choice of the interaction length and the ratio between ohmic and diffractive losses was proposed by Gregory S. Nusinovich [87].

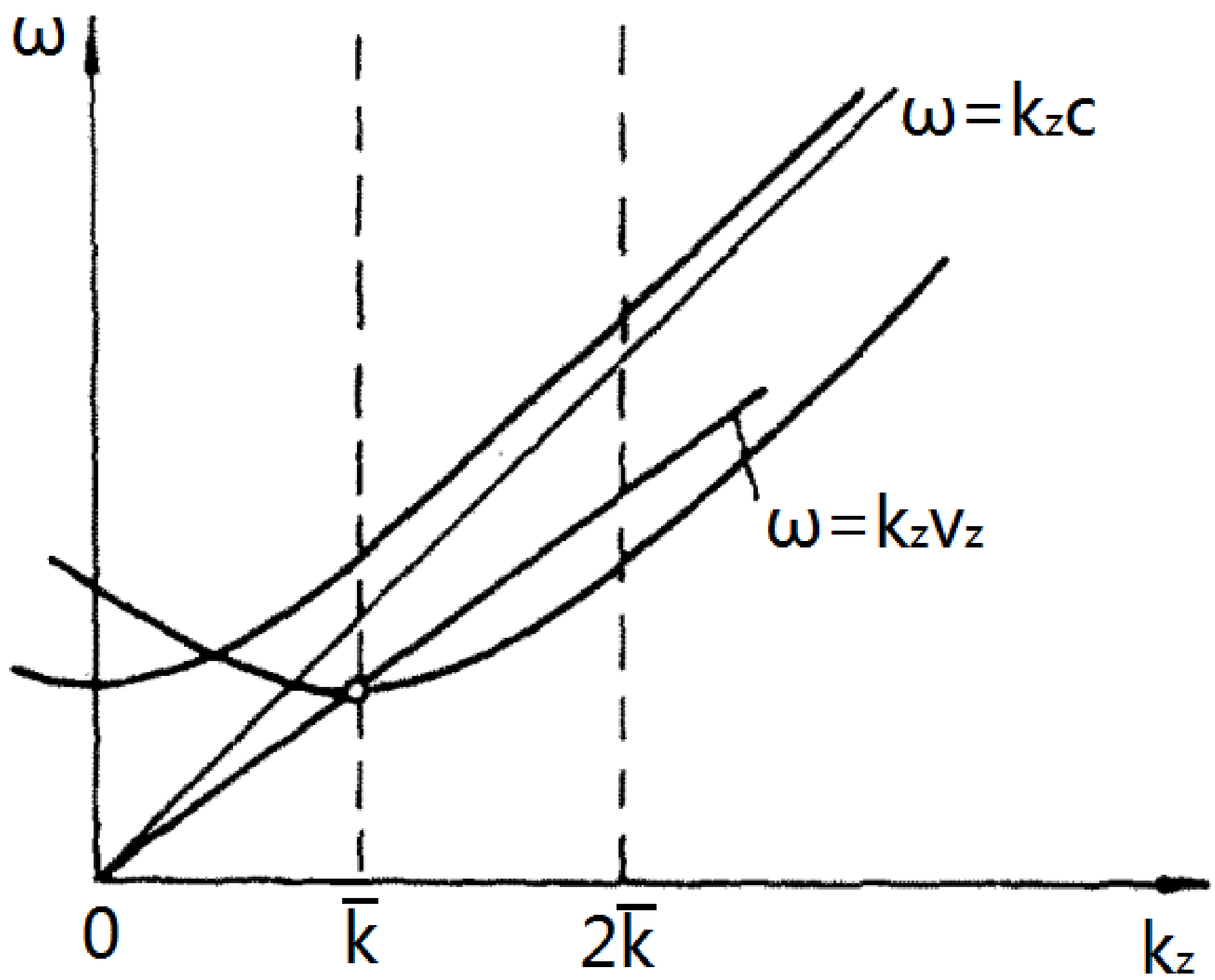

FIG. 9. The Brillouin diagram of the Orotron.

The typical schematic of the orotron is shown in Fig. 10. The ortron consists of a cathode, a collector, a flat mirror with periodic structure, a concave mirror, an output waveguide, etc. The open cavity is used to provide the effective selection of transverse modes [88]. The orotrons have the ability to produce millimeter waves and even THz (terahertz) waves. The orotrons can output power from several mW to tens of kW in weakly relativistic devices, and can output power up to hundreds of MW in strong relativistic devices. The output frequency of the orotrons ranges from 10 GHz to 361 GHz was observed in the orotron experiments [88].

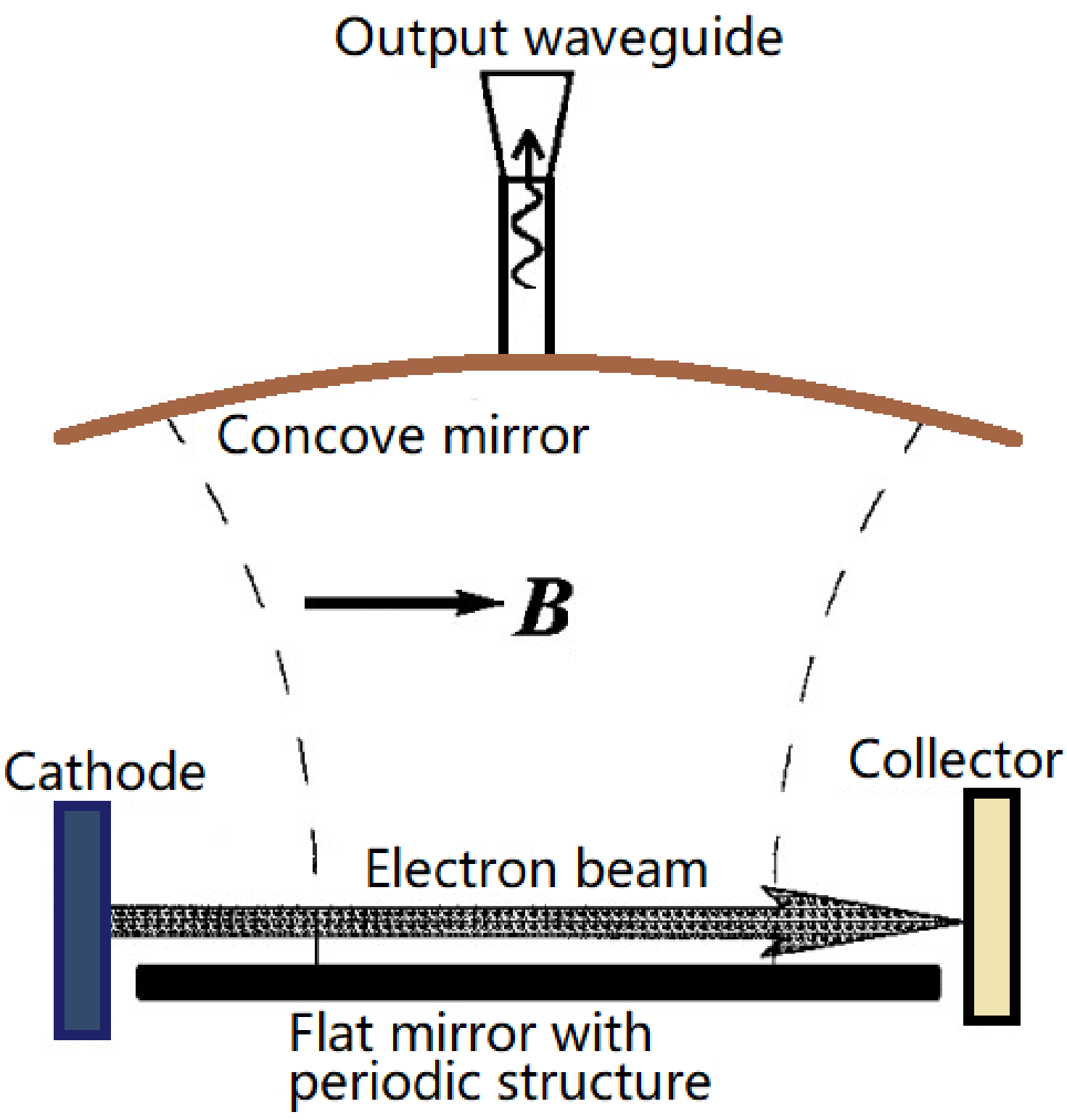

FIG. 10. The basic structure of the Orotron.

TABLE X. The progress in experimental research of Orotrons.

| Band | Peak power [W] | Freq. [GHz] | Efficiency [%] | Pulse length [s] | Institute | Years | Remark |
|---|---|---|---|---|---|---|---|
| K | 200[89] | 25 | | 1μ | IPP, USSR | 1969 | |
| V | 100m[90] | 53-73 | | 60μ | AER, HDL | 1981 | |
| mm | 50m[88] | 140 | | 10m | IAP | 2002 | |
| THz | 30m | 370 | | 10m | IAP | 2002 | |
| Ka | 120M[22] | 37.5 | 4~10 | 6n~8n | IAP | 1984 | Strong relativistic |
| V | 50M | 60 | 5 | 6n~8n | IAP | 1984 | Strong relativistic |

Note: AER, HDL is short for U.S. Army Electronics Research and Development Command, Harry Diamond Laboratories. IPP, USSR is short for Inst. for Phys. Problems, Moscow, Soviet Union.

## 2.7. Flimatron (Smith-Purcell Free Electron Maser)

If the electrons in the Cherenkov oscillator are synchronized with the first spatial harmonics of the wave, the phase velocity of the wave is close to the speed of light,

$$\omega \simeq k_z c, \tag{5}$$

then the Smith-Purcell Free Electron Maser realized [91,92]. The **fl**icker **ima**ge of the particles is different from the oscillation of the particles in a maser, so this device was called the flimatron. The radiation of a wave is similar to a free electron laser, and the angular frequency of the wave is,

$$\omega \simeq \gamma^2 \Omega, \quad (6)$$

where,

$$\Omega \simeq \bar{k} v_z. \quad (7)$$

The research progress is shown in Table XI.

TABLE XI. The progress in experimental research of Flimatrons.

| Band | Peak power [MW] | Freq. [GHz] | Efficiency [%] | Pulse length | Institute | Years |
|---|---|---|---|---|---|---|
| K | 60[22] | 25 | 5 | ~ns | IAP | 1984 |
| V | 20 | 61.22 | 5 | ~ns | IAP | 1984 |
| W | 3 | 90.91 | 5 | ~ns | IAP | 1984 |

## 2.8. Magnetron

Magnetron is a kind of orthogonal field oscillator, which can also be classified as a Cherenkov radiation device. It has the advantages of high efficiency and low cost and is widely used in radar, industrial microwave heating, household microwave ovens, and other fields. At the beginning of the 20th century, American Albert W. Hull first invented the magnetron [93, 94]. In the 1930s, H. Ф. Alekseréff of the Soviet Union and J. T. Randall of the United Kingdom and others developed the multi-cavity magnetron with practical value. The relevant history can be found in [95]. In World War II, the multi-cavity magnetrons were widely used in military radars, which greatly promoted the development of magnetrons. Subsequently, many new types of magnetrons such as coaxial magnetrons, voltage-tuned magnetrons, and long anode magnetrons were invented. Reports of relativistic magnetrons began to appear in the 1970s.

The magnetron includes an electrical power input circuit, a magnet, a cathode, a slow-wave structure (anode), and an output structure. The magnetron usually works in $\pi$ mode or $2\pi$ mode (in the case of $\pi$ mode, the phase of the microwave electric field at the mouth of two adjacent resonators is 180 ° different). The electrons emitted from the cathode make a cycloidal motion under the action of an orthogonal electromagnetic field. Adjust the DC voltage and magnetic field so that the average drift velocity of the electrons in the circumferential direction is equal to the phase velocity of the microwave field, and the electrons can interact with the microwave. The electrons in the microwave deceleration field gradually transfer the energy to the microwave field, move toward the anode, and are finally collected by the anode. These electrons transfer energy to the microwave field, which is conducive to the establishment of microwave oscillations in the magnetron, which can be called favorable electrons. Those electrons in the microwave acceleration field get energy from the microwave field and move toward the cathode, and finally hit the cathode. Electrons in the microwave acceleration field are called unfavorable electrons. The unfavorable electrons emit a large number of secondary electrons when they bombard the cathode, which can increase the number of electrons in the interaction zone. The maximum deceleration field is the clustering center of electrons, and the electrons on both sides of it move toward the clustering center. The maximum acceleration field area is the center of the electron's divergence, and nearby electrons move to the left and right sides, and finally turn into favorable electrons. In this way, during the establishment of the oscillation, the number of unfavorable electrons is decreasing, the number of favorable electrons is increasing, and they are concentrated toward the cluster center, and a spoke-shaped electron cloud is gradually formed in the interaction space, as shown in Fig. 11. As the microwave field in the interaction zone decays exponentially away from the anode surface, the microwave field on the cathode surface is very weak, and the clustering of electrons is extremely small. There will be no obvious electronic spokes near the cathode, but a uniform distributed electronic wheel. The overall effect of the interaction between electrons and microwaves in a magnetron is that the electrons give energy to the microwave field, establishing a stable microwave oscillation in the magnetron.

Since the advent of magnetrons, a large number of magnetron theories have been proposed [96-98], which greatly helped the development of magnetrons. With the development of computer technology, the PIC program represented by MAGIC is widely used in the design and simulation of magnetrons [99].

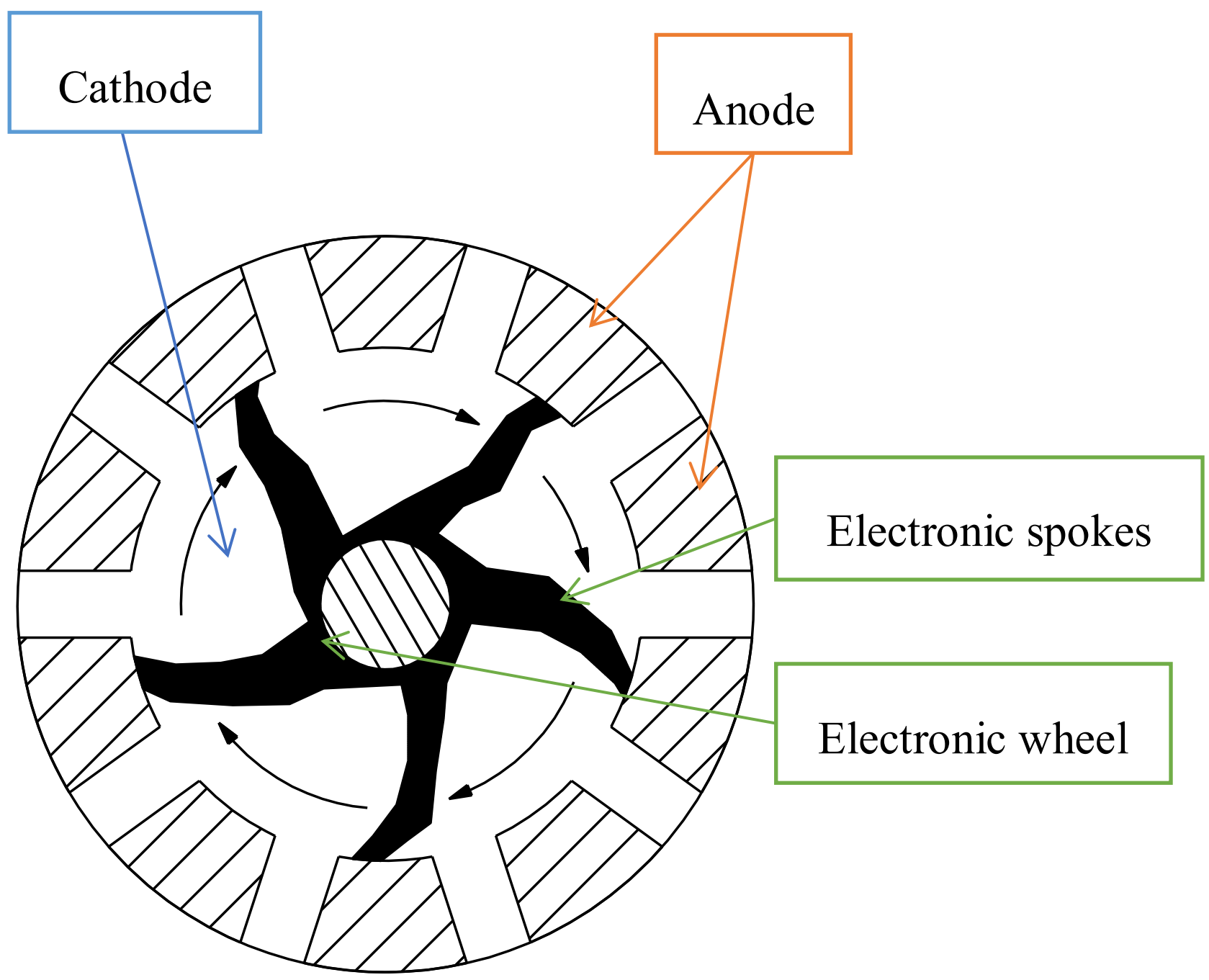

FIG. 11. The schematic diagram of the electronic spokes of a magnetron.

The efficiency of the magnetron can reach 80%. At the frequency of 915 MHz, the continuous wave output power can reach 100 kW; at the frequency of 2.45 GHz, the continuous wave output power can reach 30 kW [100]. In terms of bandwidth, although orthogonal field devices like a magnetron are not as good as the traveling wave tubes, they are better than the klystrons. In the field of magnetrons for microwave ovens, the main application frequency is 2.45 GHz with an efficiency of 70% [101]; the frequency of 5.8 GHz can also be used in microwave ovens, with efficiency currently around 50% and power close to 1 kW [102,103].

In the field of military radar, after several generations of development, at the X-band, with a peak output power of 25kW, the life of the magnetron can reach 6000+ hours; when the output peak power is 4kW, the life can reach 15k + hours. The latest generation of military magnetrons has further improved its performance by improving mode stability, developing new tuned output systems, and adopting new rare-earth magnets [104].

The use of non-$\pi$ mode spatial harmonic magnetrons is also an attractive option for developing high-power millimeter wave, terahertz sources. For example, N. I. Avtomonov of the Institute of Radio Astronomy of the National Academy of Sciences of Ukraine (IRANASU) reported that a space harmonic magnetron can output a peak power of 1.3 kW and an average power of 0.13 W at a frequency of 210 GHz. It can continuously run for 500 h with a 50 ns pulse width and a 0.01% duty cycle [105]. They have developed a series of millimeter wave space harmonic magnetrons, with frequencies from 95 GHz to 225 GHz, and the peak power up to 20 kW at 95 GHz [106].

Part of the research progress of magnetrons is shown in Table XII. Among them, CPI has produced a series of magnetrons for air traffic control (frequency 2.7GHz-34.5GHz, peak power up to 1MW), weather radar (frequency 2.7GHz-8.5GHz, peak power up to 1MW), frequency agile radar (frequency 5.45GHz-32.1GHz, peak power up to 800kW), industrial applications (896MHz-928MHz, continuous wave power 30-125kW), etc. At present, the main research directions include high power, high frequency, high efficiency, high pulse width or high repetition frequency, miniaturization, phase locking [107], and tunability.

TABLE XII. The progress in experimental research of non-relativistic magnetrons.

| **Band** | **Peak power [W]** | **Average power [W]** | **Freq. [Hz]** | **Efficiency [%]** | **Pulse length [s]** | **Duty cycle [%]** | **Beam voltage [V]** | **Beam current[ A]** | **Institute** | **Model** |
|---|---|---|---|---|---|---|---|---|---|---|
| S | 19k[107] | 19k | 2.45G | 71 | CW | | 11.7k | 2.3k | Sichuan University, China | |
| C | 2.5M | 2.5k | 5.7G±10MH | 45 | 4μ | 0.1 | 45-50k | 110 | CPI | VMC3109 |
| C | 700[108] | | 5.8G | 50 | | | 4.7k | 0.3 | UESTC | |
| C | 1k[103] | | 5.8G | 58 | 10ms (5Hz) | | 4.34k | 0.416 | CAEP | |
| X | 1.5M | 2.7k | 9.3G±30MHz | 45 | 3.5μ | 0.18 | 34-37k | 90A | CPI | VMX3100HP |
| W | 20k | | 95G | | 0.1μ | | | | IRANASU | |
| mm | 1.7k | | 225G | | 0.05μ | | | | IRANASU | |

Note: UESTC is short for University of Electronic Science and Technology of China.

In the 1970s, G. Bekefi, T. J. Orzechowski and other scientists began to apply pulse power technology and explosion-cathode technology to magnetrons, and developed relativistic magnetrons [109,110]. The main R&D institutions of relativistic magnetrons include PI (Physics International Company, United States), TINP (Tomsk Institute of Nuclear Physics, Russia), IAP (Institute of Applied Physics, Russia), UESTC (University of Electronic Science and Technology of China), and so on. The frequency of the microwave generated by the relativistic magnetron has covered the range of 1 ~ 10 GHz, and the microwave output power of a single device reaches several GW. Some research progress is shown in Table XIII.

TABLE XIII. The progress in experimental research of relativistic magnetrons.

| **Band** | **Peak power [W]** | **Freq. [GHz]** | **Efficiency [%]** | **Pulse length [s]** | **Beam voltage [V]** | **Beam current [A]** | **Institute** | **Remark** |
|---|---|---|---|---|---|---|---|---|
| S | 540M[111] | 2.68 | 6.5 | ~40n | 489k | 16.9k | UESTC | Permanent magnet packaging |
| S | 1G~1.6G[112] (one-vane extraction) | 2.83-2.95 | 6-10 | <35n | 750k | 21k | PI | |
| S | 2.4G-3.6G (six-vane extraction) | 2.85-2.90 | 15-23 | <35n | 750k | 21k | PI | |
| S | 1.7G[110] | 3 | 35 | 30n | 360k | 12k | MIT | |
| C | 6.9G[113] (six-vane | 4.5 | 9 | 20n-40n | 1.2M | 10k | PI | |

| | extraction) | | | | | | | |
|---|---|---|---|---|---|---|---|---|
| L | 400M[114] | 1.21 | 10 | 70n | | | PI | Frequency tunable 23.9% |
| **S** | 500M | 2.82 | 10 | 70n | | | PI | Frequency tunable 33.4% |
| **S** | ~1G[111] | 2.78 | ~10 | | ~620k | ~16.4 | UESTC | Frequency tunable 18% (500MHz) |

Note: MIT is short for Massachusetts Institute of Technology.

## 2.9. Crossed-Field Amplifier (CFA)

Crossed-field amplifier (CFA) is another type of orthogonal field device developed on the basis of magnetrons. It began to develop in the middle of the 20th century. CFAs are widely used in radar transmitters and are generally used as the last stage of the amplification link. The CFA has the advantages of high efficiency, low operating voltage, high phase stability, compact size, etc., and its peak power can reach several MW [115]. CFAs can use cold cathodes without preheating and can be instantaneously started. The main disadvantages of CFAs are low gain and large noise. The gain is generally only 10~20 dB. Improving gain is the main research direction of CFAs.

The cross-field amplifier is generally composed of a magnet, an electron gun, a slow-wave structure (anode), a bottom electrode, a collector, and a microwave output structure. According to whether the direction of the phase velocity and the group velocity of the waves are in the same direction, the CFA can be divided into two types: forward wave CFA and backward wave CFA [116]. According to the electron beam injection method, it can be divided into injection type and distributed emission type. The distributed emission device has only one cathode (i.e., the electron gun). The injection type device has a bottom electrode in addition to an electron gun. According to the structure, the CFAs can be divided into reentrant CFA and non-reentrant CFA. Reentrant devices do not have separate collectors. The anode collects the remaining electrons. At present, the most widely used CFA device is the reentrant, distributed-emission, forward wave CFA. CFAs usually have slow wave structures similar to TWTs. The forward wave CFA usually uses helix while backward wave CFA usually uses the bar line. The structure of a typical reentrant, distributed-emission, forward wave CFA and a non-reentrant, injection-type, forward wave CFA are shown in Fig. 12. The progress in experimental research of CFAs is shown in Table XIV.

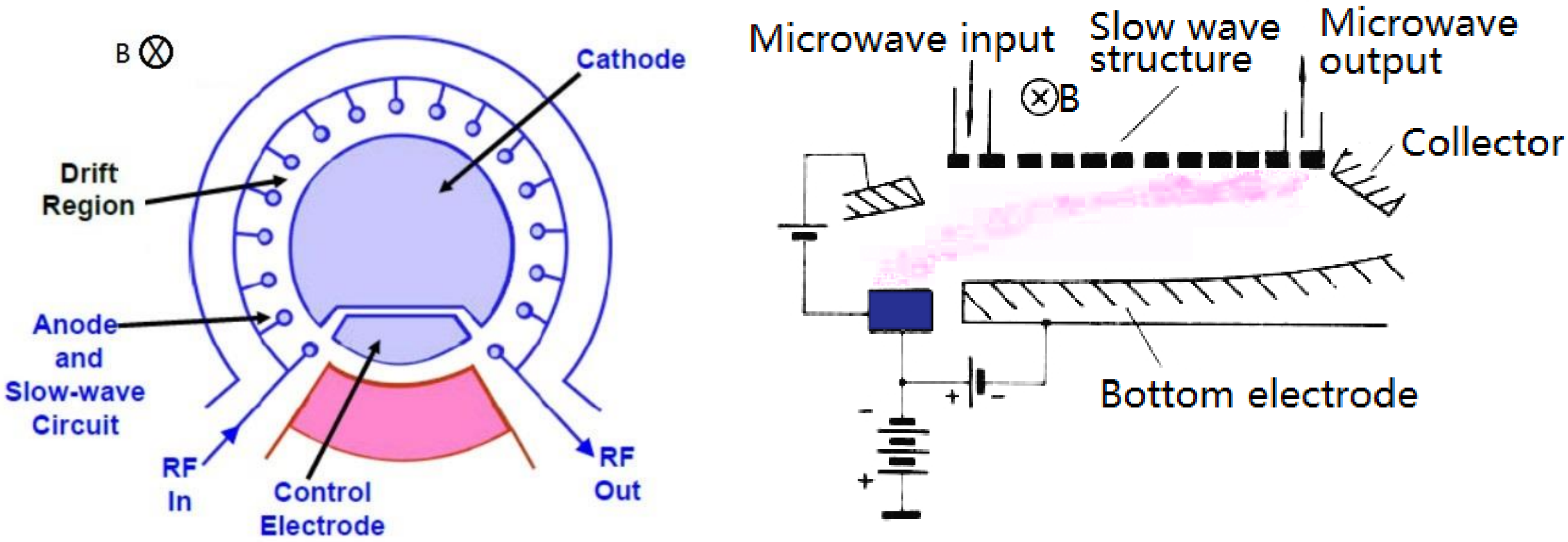

FIG. 12. The simple structure of the CFAs. (a) Reentrant, distributed-emission, forward wave CFA; (b) Non-reentrant, injection-type, forward wave CFA.

TABLE XIV. The progress in experimental research of CFAs.

| Band | Peak power [W] | Average power [W] | Freq. [Hz] | Efficiency [%] | Gain [dB] | Pulse length [s] | Duty cycle [%] | Voltage [V] | Peak current [A] | Institute | Model |
|---|---|---|---|---|---|---|---|---|---|---|---|
| L | 90k | 2.88k | 1.3G±50M | 33 | | 40μ | 3.2 | 11k | 25 | CPI | VXL1169 |
| C | 900k | 4.5k | 5.65G±250M | | | 50μ | 0.5 | 30k | | CPI | VXC1659 |
| S | 250k[115] | 20k | 2.9G±150M | >52 | 13 | 300μ | 8 | 18k | 23.8 | CETC 12th institute | |
| S | 220k | 4.4k | 3.3G±200M | 38 | | 50μ | 2 | 16k | 36 | CPI | VXS1925 |
| X | 300k[117] | 5.1k | 9G±500M | >57 | >14 | 100μ | 1.7 | 24k | 21.8 | CETC 12th institute | |
| X | 900k | 900 | 9.2G±300M | 39 | | 0.83μ | 0.1 | 38k | 60 | CPI | SFD233G |

## 2.10. Magnetically Insulated Transmission Line Oscillator (MILO)

Magnetically insulated transmission line oscillator (MILO) and its linear theory were proposed by Raymond W. Lemke and M. Collins Clark in the United States in 1987 [118]. Its structure is shown in Fig. 13. In 1995, Steve E. Calico et al. proposed a load-limiting MILO with an anode electron collection structure [119], and obtained a peak output power of up to 1GW. Its structure is shown in Fig. 14. It consists of a cathode, a slow wave structure, an RF (radio frequency) choke structure, and a collector (beam dump). For the history of MILO, please refer to the paper [120]. Xiaoping Zhang of China's National University of Defense Technology proposed a new MILO (V-MILO) with a virtual cathode oscillator (Vircator) as the load. Under the condition of 540 kV beam voltage and 42 kA beam current, it can get a peak output power of 500 MW at 5 GHz, with an efficiency of 2.3% [120]. At present, MILO has become one of the highest single pulse ratio microwave energy output devices among all HPM devices. Various forms of MILO have been developed, including load-limiting MILO (the load types include Vircator, MILO, Axial Dode, TTO, etc.).

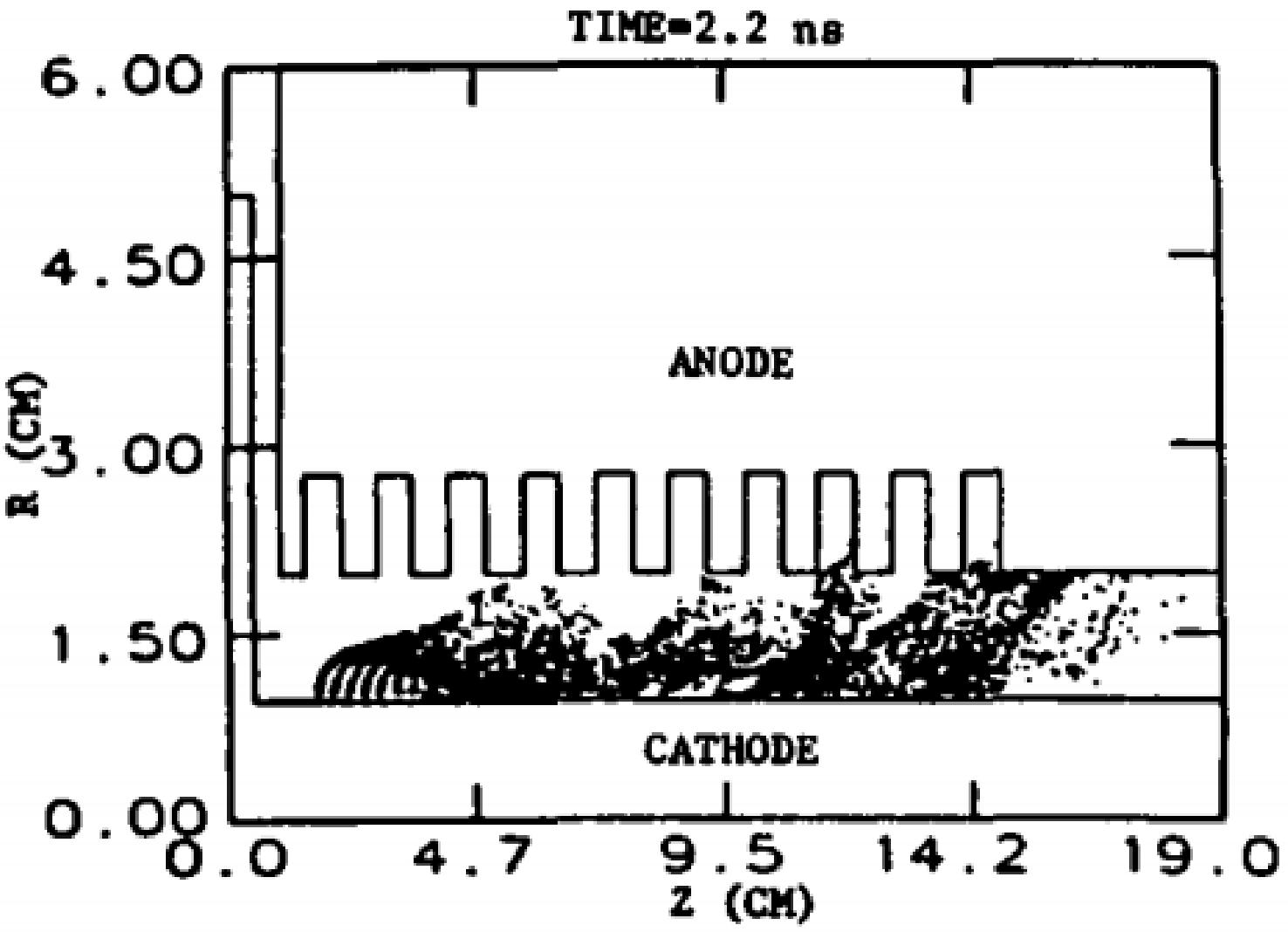

FIG. 13. The structure of the first MILO (line-limiting type).

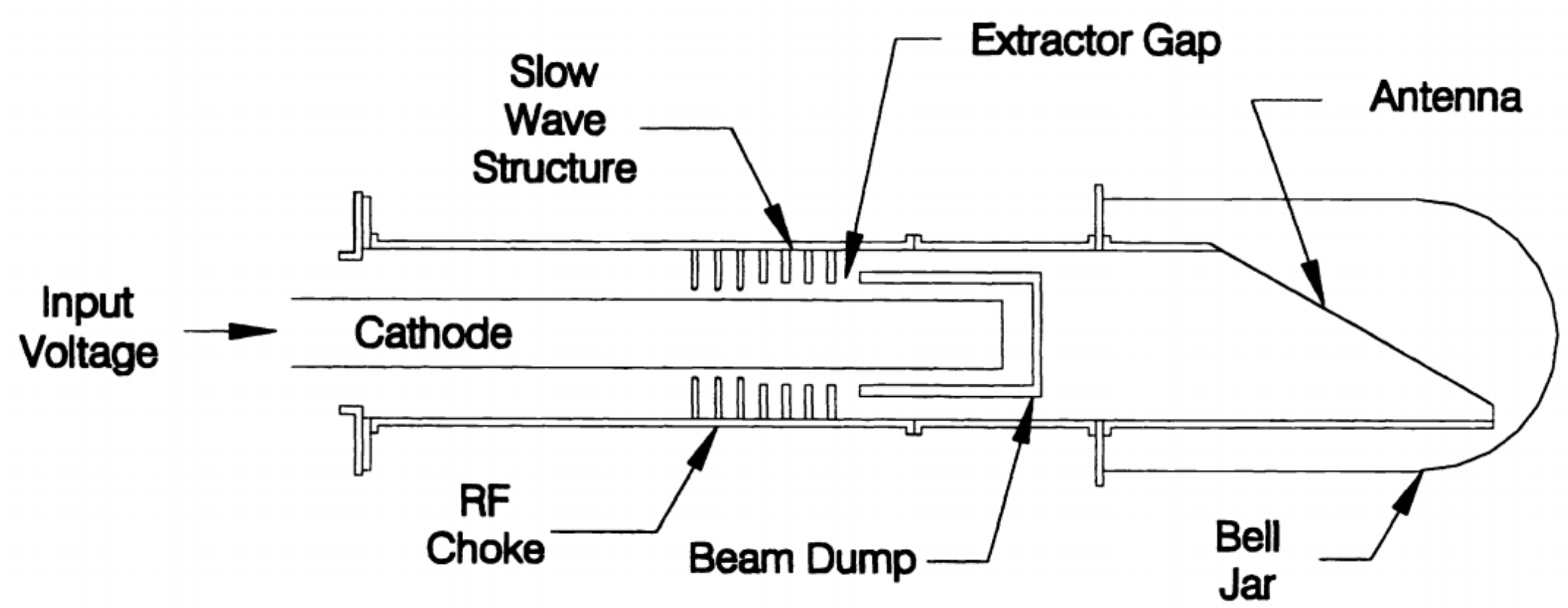

FIG. 14. The structure of the first load-limiting MILO.

The MILOs do not require an external magnetic field. The DC magnetic field is provided by the internal current through the tube. This DC magnetic field together with the orthogonal DC electric field determines the electron drift speed. The DC magnetic field generated by MILO itself can suppress the emission of electrons from the cathode to the anode. This self-insulation mechanism and low impedance can prevent the electron strike between the cathode and anode, so MILO can withstand very high power. The operation of MILO can be divided into three stages, which are magnetic insulation formation, RF growing, and microwave saturation [121], as shown in Fig. 15.

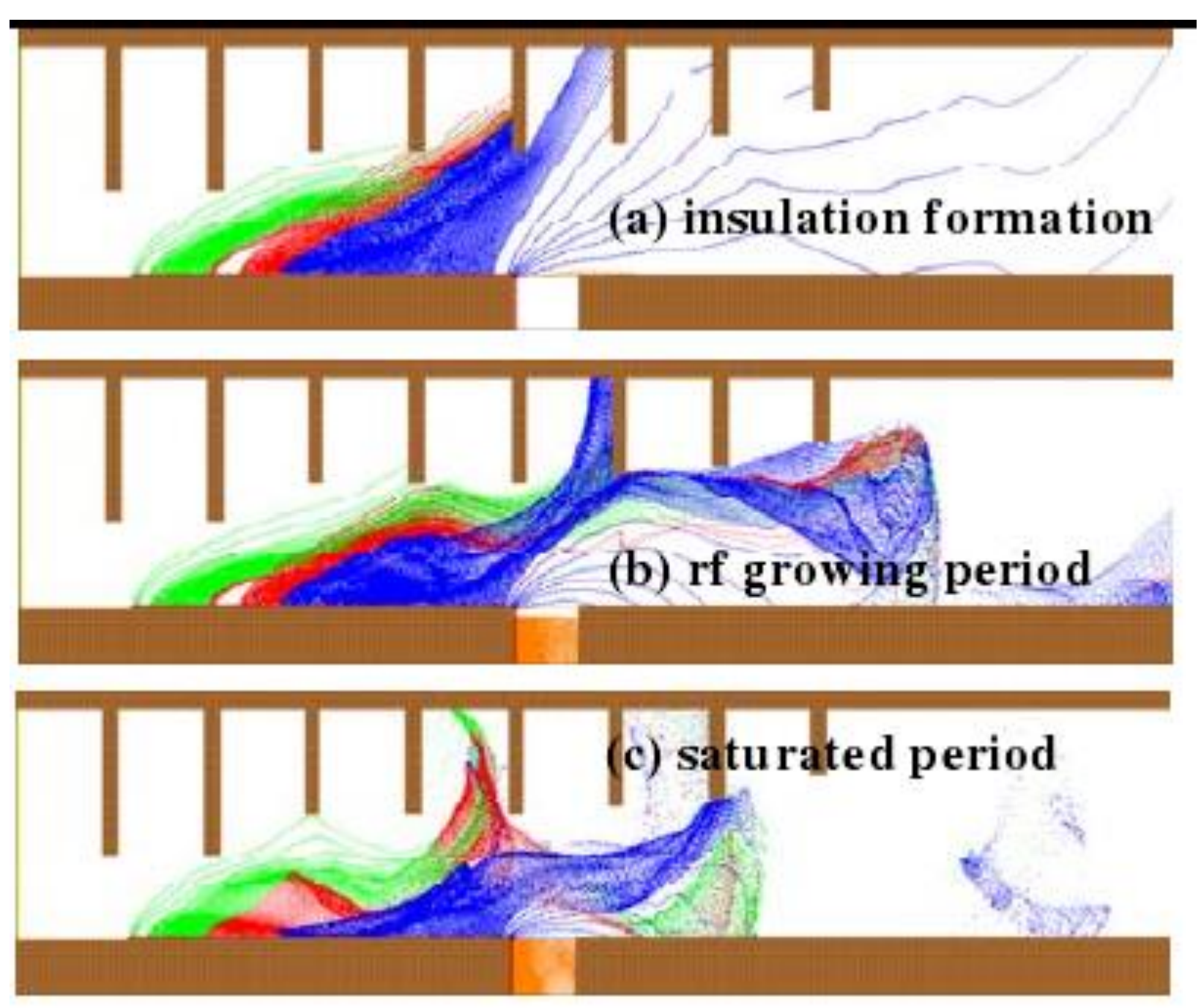

FIG. 15. Three stages of MILO operation.

Partial research progress of MILO is shown in Table XV. The main disadvantage of MILO is that the efficiency is not high. The research directions of MILO mainly include improving power, pulse width, and efficiency, solving the problem of RF breakdown, and studying the pulse shortening mechanism. For example, Michael D. Haworth of the United States proposed a field-forming cathode structure with a pulse width of 400 ns at a working voltage of 300 kV [122]. S.E. Calico of the United States designed a velvet cathode and conducted repeated frequency operation experiments. Under the voltage of 400kV, 10 pulses were run at 5Hz frequency, and the consistency did not change much [123].

TABLE XV. The progress in experimental research of MILOs.

| **Band** | **Power [W]** | **Freq. [GHz]** | **Efficiency [%]** | **Pulse length [s]** | **Beam voltage [V]** | **Beam current [A]** | **Institute** | **Type** |
|---|---|---|---|---|---|---|---|---|
| L | 2G[124] | 1.2 | 7 | 175n | 475k | 60k | U.S. Air Force Lab | Load-limiting type |
| L | 3.57G[125] | 1.23 | 7.9 | 46n | 740k | 61k | CAEP | Double step cathode & Load-limiting type |
| L | 2.2-2.5G[126] | 1.76-1.78 | 7.3-7.9 | | 515-538kV | 58-61k | NUDT | Load-limiting type |
| S | 0.9G[127] | 2.4 | 6 | 23n | 500k | 30k | CNRS | |
| S | 1G[121] | 2.59 | 4.6 | ~40n | 665k | 32.3k | NUDT | Cone-shape MILO |
| S | 500M[128] | ~2.64 | 5.7 | ~90n | 350k | 25k | NUDT | Cone-shape MILO |
| Ku | 89M[129] | 12.9 | 0.3 | 15n | 539k | 57k | CAEP | |

Note: CNRS is short for French National Centre for Scientific Research.

# 3. Transition Radiation Devices

When electrons cross the boundary between two media with different refractive indices, or through some perturbation in the media, such as a conductive grid or a gap on the surface of a conductor, transit radiation occurs. The main difference between transit radiation and Cherenkov radiation is that the field interacting with the electron beam is a standing wave field. There are fewer types of transit radiation devices, including klystrons, transit time oscillators (TTO), etc.

## 3.1. Klystron

A klystron is a microwave vacuum tube that uses periodic modulation of the electron beam velocity to achieve oscillation or amplification. It first modulates the velocity of the electron beam in the input cavity (buncher cavity), converts it into density modulation after drifting, and then exchanges the energy of the clustered electrons with the microwave field to complete microwave oscillation or amplification [130]. The basic structure of the klystron is shown in Fig. 16, which is composed of the cathode (electron gun), resonant cavity, magnet, collector, and so on.

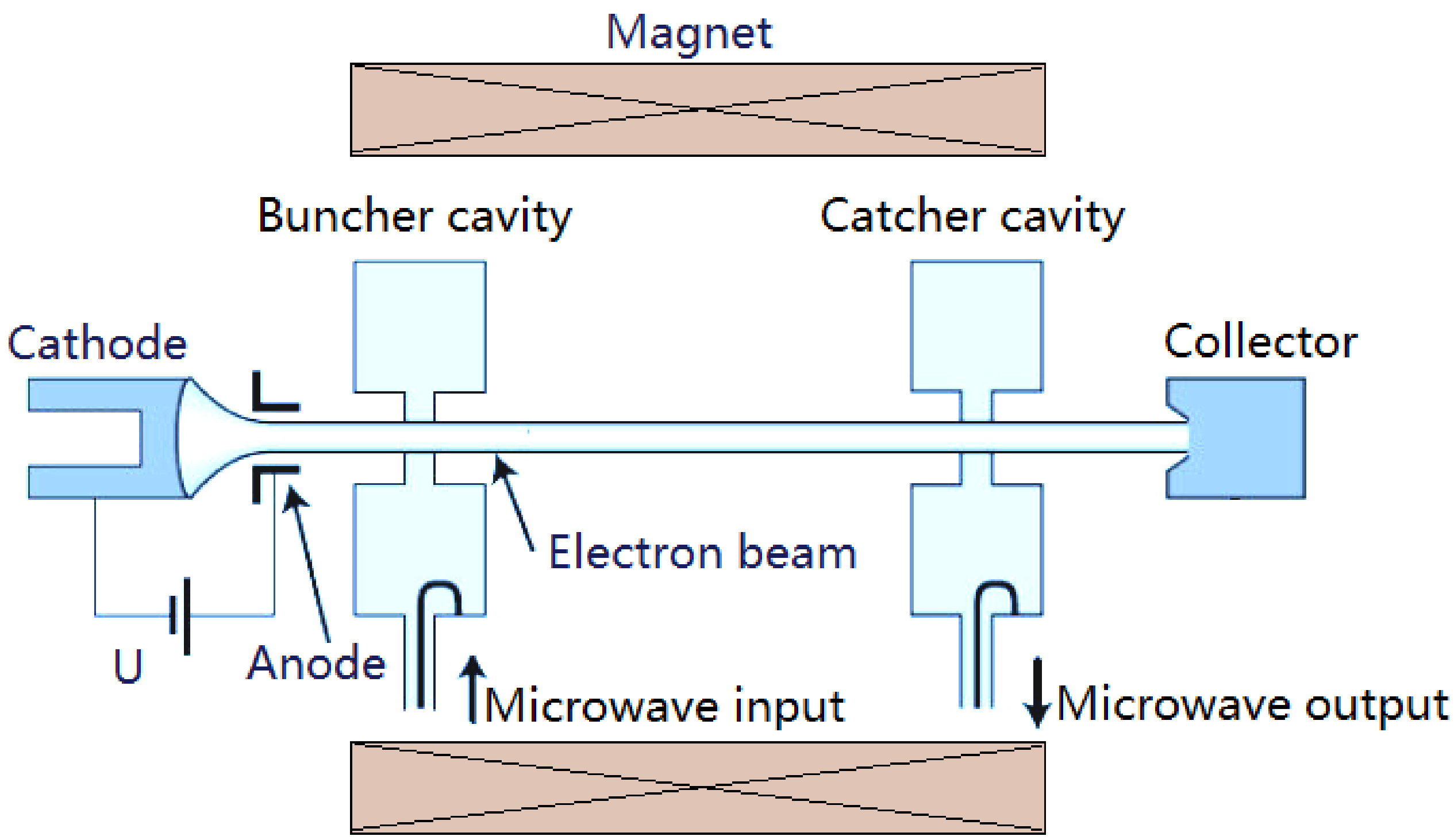

FIG. 16. The structure of klystron.

Since the 1930s, D. A. Rozhansky and others in the former Soviet Union began to notice the phenomenon of density modulation after the electron beam received speed modulation. In the late 1930s, the American Varian brothers invented the first double-cavity klystron. During the development of the klystron, various types such as Multi-Beam Klystron (MBK), Sheet Beam Klystron (SBK), and Extended Interaction Klystron (EIK) were produced.

In the early 1940s, V. F. Kovalenko of the former Soviet Union and J. Bernier of France proposed the concept of multi-beam klystron. In the early 1960s, S. Koroljev of the former Soviet Union developed a practical multi-beam klystron [131]. The R&D institutes of MBK mainly include ISTOK, TORIY, the Naval Research Laboratory (NRL) of the United States, etc. [132]. The main characteristics of multi-beam klystrons are low operating voltage, wide frequency band, small size, etc., which can achieve a wide frequency band at lower peak power.

The concept of a band-shaped klystron was first proposed by Kovalenko and others in the former Soviet Union in the 1930s. Due to the difficulty of analysis of non-axisymmetric structures, no substantial progress was made until the beginning of the 21st century. The research institutions of SBK mainly include Stanford Linear Accelerator Center (SLAC) [133], Calabazas Creek Research, Inc. (CCR, United Stated), Technische Universität Berlin (TU-Berlin, Germany), etc. [134]. CPI has developed an X-band SBK with a peak power of 2.67 MW [135], and the IECAS (Institute of Electronics, Chinese Academy of Sciences) has also developed an X-band SBK with a peak power of 2.8MW [136].

In the 1960s, M. Chodorow and T. Wessel-Berg of the United States first proposed an extended interaction klystron [137], which combined the advantages of a klystron and a coupled cavity traveling wave tube, and had higher gain and bandwidth than traditional klystrons. The EIK has two types: amplifying tube and oscillating tube. It is especially suitable for working in the millimeter wave and terahertz bands. EIK's main research units include the CPI and NRL. The frequency of the EIK produced by CPI is from 25GHz to 700GHz, and the maximum continuous wave power can reach 1kW.

Klystron has the advantages of high power, high gain, and high stability. Klystrons are widely used, of which continuous wave klystrons are mainly used in synchrotrons, magnetic confinement fusion, industrial microwave heating, etc. In the P and L bands, continuous wave power of klystrons can range from hundreds of kW to MW; in the C band, continuous wave power can reach hundreds of kW. By reducing the conductivity of the electron beam and the use of second harmonic resonant cavities, the efficiency is close to 70%.

The application of klystrons in early warning and guided radar requires a wide frequency band. Through the use of overlapping mode double-gap coupling cavity, filter loading and other technical means, the instantaneous working bandwidth of klystrons has been greatly expanded. At the MW level, the bandwidth of L and S-band klystrons has reached 10% [138]. Some progress of klystron research is shown in Table XVI.

TABLE XVI. The progress in experimental research of non-relativistic klystrons.

| **Band** | **Peak power [W]** | **Average power [W]** | **Center freq. [GHz]** | **1dB bandwidth [MHz]** | **Gain [dB]** | **Efficiency [%]** | **Pulse length [s]** | **Duty cycle [%]** | **Institute** | **Model** |
|---|---|---|---|---|---|---|---|---|---|---|
| P | 1.1M[139] | 1.1M | 0.352 | | 40 | 68 | CW | | TED | TH2089 |
| P | 80k | 80k | 0.5 | | 40 | 60 | CW | | CPI | VKP-7957A |
| P | 1M | 1M | 0.7 | ±0.7 | 40 | 65 | CW | | CPI | VKP-7952A |
| P | 2.5M | 250k | 0.805 | ±0.7 | 45 | 55 | 1.67m | 10 | CPI | VKP-8290A |
| L | 1.2M | 1.2M | 1.25 | | | 63 | CW | | TOSHIBA | E3718 |
| L | 10M | 150k | 1.3 | | 49 | 66 | 1.5m | | TOSHIBA | E3736 (MBK) |
| L | 1M | 100k | | | 50 | 40 | | | IECAS | KL4001 |
| S | 0.7-1.2M | 11-20k | 2 | ±120 | | | | | ISTOK | (MBK) |
| S | 6M | 25k | 2.45 | | | | | | TORIY | (MBK) |
| S | 1.2M | | 3.3 | ±200 | | | | | CPI | VKS8340 |
| C | 250k | 250k | 4.6 | | 56.3 | 44.3 | CW | | CPI | VKC-7849 |
| C | 200k | 10k | 5.6 | ±126 | 45 | 35 | | | IECAS | KC4079D (MBK) |
| C | 20k[140] | 20k | 7.17 | ±22.5 | | | CW | | NASA | |
| X | 100k | 100k | 7.19 | ±45 | | | CW | | NASA | |
| X | 20k[141] | 20k | 7.19 | ±45 | | | CW | | Rheinmetall (RhI) | |
| X | 250k | 250k | 8.56 | ±10 | 45 | 41 | CW | | CPI | VKX-7864B |
| X | 250k | 250k | 8.51 | ±10 | 53 | 50 | CW | | CPI | VKX-7864A |
| X | 50k | 1.2-2k | 9.75 | ±250 | 40 | 30 | | | TMD | PT6203 |
| X | 2.67M | | | | 45 | 45 | | | CPI | (SBK) |
| Ka | 1k | 1k | 27.5-31 | ±125 | 40 | | CW | | CPI | VKA2400 S20 (EIK) |
| Ka | 800 | 800 | 34-36 | ±250 | 46 | | CW | | CPI | VKQ2488 (EIK) |
| Ka | 3k | 300 | 34-36 | ±100 (-3dB bandwidth) | 45 | | | 10 | CPI | VKQ2470 (EIK) |
| W | 2k[142] | 28 | 94.05 | ±125 | 56 | 32 | | 1.4 | CPI | VKB2469 T16（EIK） |
| mm | 20 | 20 | 110-140 | ±100 | 30 | | CW | | CPI | VKT2480 （EIK） |
| mm | 9 | 9 | 170-220 | ±100 | 23 | | CW | | CPI | VKY2444 （EIK） |

Increasing the current and voltage of the electron beam to cause a relativistic effect is a relativistic klystron, which is mainly used in high-energy accelerators, medical accelerators, and other fields. The use of a ring electron beam can reduce the space charge effect. A strong electron relativistic klystron of a ring electron beam was studied by M. Friedman of NRL in the 1980s. A peak power output of 15GW was obtained in the L-band and a peak power output of 1.7GW was obtained in the S-band. However, due to insufficient understanding of non-linear phenomena such as virtual cathode formation and electron reflection, it is difficult to improve efficiency [143]. The China Academy of Engineering Physics (CAEP) started research on the relativistic klystron from the 1990s and currently has achieved a peak power output of more than 1GW [144,145]. Japan's Toshiba used a solid single-beam klystron to achieve a peak power output of 74MW in the X-band. It is generally called a weak-flow relativistic klystron. Some typical progress of the relativistic klystron is shown in Table XVII. The development direction of klystron mainly includes further increasing power and frequency, increasing bandwidth, improving efficiency, and increasing service life.

TABLE XVII. The progress in experimental research of relativistic klystrons.

| **Band** | **Peak power [W]** | **Average power [W]** | **Center freq. [GHz]** | **Gain [dB]** | **Efficiency [%]** | **Pulse length [s]** | **Voltage [V]** | **Current [A]** | **Institute** | **Model** |
|---|---|---|---|---|---|---|---|---|---|---|
| L | 15G | | 1.3 | | 40 | | | | NRL | |
| S | 200M | | 2.856 | 58 | 47 | 1μ | 610k | 780 | SLAC | |
| S | 1.2G | 2.4k | 2.95 | 34 | 24 | 20n | 700k | 6.5k | CAEP | |
| S | 1.7G | | 3.5 | | 60 | <80n | | | NRL | |
| C | 50M | 6.25k | 5.712 | 52 | 47 | 2.5μ | 354k | 315 | TOSHIBA | E3746A |
| X | 75M | 14-29k | 11.424 | | 55 | 3.2μ | 490k | 257 | SLAC | XP3 |
| X | 74M | 16.6k | 11.424 | 60 | 55 | 1.5μ | 500k | 270 | TOSHIBA | E3768 |

## 3.2. Transit Time Oscillator

The transit time oscillator (TTO) originates from a monotron [146], as shown in Fig. 17. In a monotron, the resonant cavity must be long enough to allow the electron transit time to exceed the oscillation period, resulting in a longer resonant cavity length. When a high-current electron beam passes through a long resonant cavity, the contraction effect related to beam transmission, the expansion effect of space charge, and the reflection of electrons make the electron beam have a large energy divergence when it enters the extraction cavity. The efficiency is very low and the monotron has never been practically used.

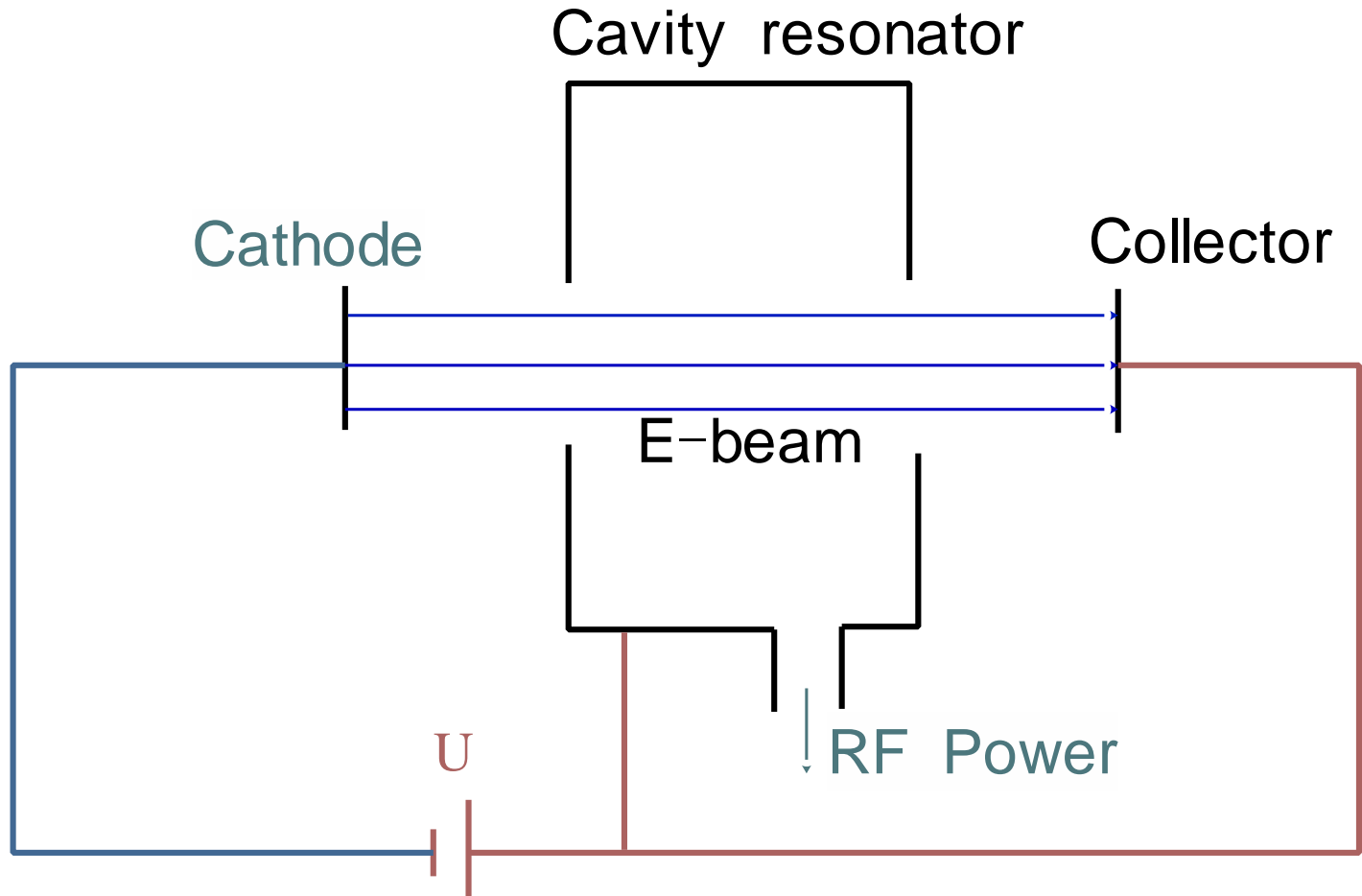

FIG. 17. The simple structure of monotron.

Until the early 1990s, Barry M. Marder et al. proposed the use of a split-cavity oscillator (SCO) to modulate high-current relativistic electron beams. A metal foil through which the electron beam can pass is added in the cavity of the monotron. This metal foil does not completely separate the cavity, and the electromagnetic field between the divided cavities can be coupled through the gap between the foil and the cavity wall, which is shown in Fig. FIG. 18.

The characteristic of this structure is that it has a unique mode. In this mode, the amplitude of the field in the two cavities is the same and the phase is 180 degrees different.

In this way, it realizes the change in the field required for the unstable interaction with the electron beam 'in space', but not 'in time' like a monotron. Therefore, a longer transit time is not required, which makes the device more compact and also solves the problem of hooping of strong current beams and space charge effects caused by excessive transit times. Studies have shown that when the cavity length of the SCO is a quarter of the wavelength, the efficiency of converting the electron beam energy into the microwave energy is the highest. Barry et al. used a beam voltage of about 150 kV and a beam current of about 2 kA., the experimental microwave output power was about 25 MW, the pulse width was about 800 ns, and the beam-wave conversion efficiency was about 8% at a frequency of 1.3 GHz [147,148]. Before entering the extraction cavity, the energy divergence of the electron beam should be minimized so that as many electrons in the electron beam have the same transit time in the extraction cavity as possible to obtain the maximum beam-wave conversion efficiency. The SCO can make the electron beam go through a period of drift to reduce the energy spread of the beam electrons and improve the efficiency.

R. B. Miller and others proposed Super-Reltron [149] in the early 1990s, which is also a type of TTO. After leaving the resonant cavity, the electron beam of Super-Reltron is accelerated again to reduce the energy divergence. At the same time, the acceleration also increases the total energy of the electron beam, which can increase the output microwave power. It achieved a peak power of 600 MW and an efficiency of 40% at a frequency of 1 GHz; it achieves a peak power of 350 MW and an efficiency of 50% at a frequency of 3 GHz [150]. J. J. Barrosoc from Associated Plasma Laboratory, National Institute for Space Research (INPE), Brazil; Zhikai, Fan and Qingxiang

Liu from the Chinese Academy of Engineering Physics (proposed the three-cavity TTO); Yibing Cao from the National University of Defense Technology (proposed the new no foil, coaxial TTO [151]) have also done a lot of researches on TTOs and made some progress, as shown in Table XVIII.

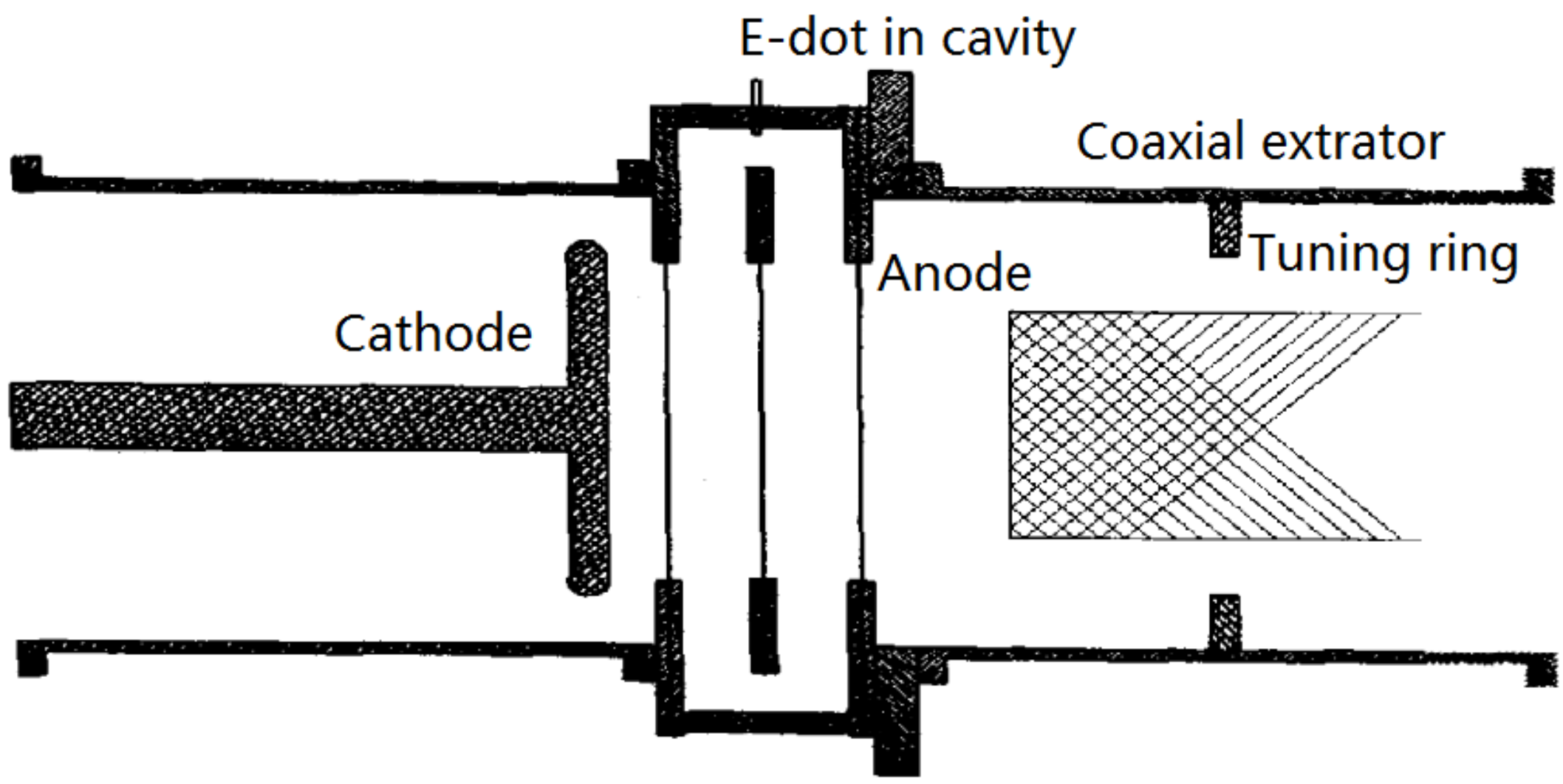

FIG. 18. Shematic of SCO with coaxial energy extractor.

TABLE XVIII. The research progress of TTOs.

| **Band** | **Peak power [W]** | **Center freq. [GHz]** | **Efficiency [%]** | **Pulse length [s]** | **Voltage [V]** | **Current [A]** | **Institute** | **Remark** |
|---|---|---|---|---|---|---|---|---|
| L | 25M | 1.3 | 8 | 800n | 150k | 2k | SNL | |
| L | 600M | 1.04 | 40 | 300n | 850k | 1.35k | TAIT | |
| S | 350M | 3.04 | 50 | 100n | 600k | 1k | TAIT | |
| S | 400M[152] | 3.8 | 17 | 15n | 521k | 5k | CAEP | The author believes that 3.8GHz belongs to the C-band. In fact, according to the IEEE standard, it should belong to the S-band. |
| C | 100k[153] | 5.74 | 7.5 | | 33k | 40 | APL, NISR | Simulation |
| C | 219.8M[154] | ~5 | 16.6 | | | | NUDT | |
| X | 2.7G[155] | 9.38 | 26.2 | | 710k | 14.5k | NUDT | Simulation |

Note: SNL is short for Sandia National Laboratories, United States. TAIT is short for Titan Advanced Innovative Technologies, United States. APL, NISR is short for Associated Plasma Laboratory, National Institute for Space Research (INPE), Brazil.

# 4. Bremsstrahlung Devices

Bremsstrahlung refers to the radiation when electrons move in an external electromagnetic field at a varying speed. Generally speaking, electrons make the oscillating motion. The frequency of the electromagnetic wave radiated by the electron is consistent with the frequency of its oscillation, or the frequency of a certain harmonic that it oscillates. Microwave sources based on this include gyrotrons, free electron lasers, virtual cathode oscillators, and so on.

## 4.1. Gyrotron

The concept of electron cyclotron can be traced back to the 1920s [156], and made important developments in the 1950s. Australian astronomer R. Q. Twiss proposed the concept of electron cyclotron resonance stimulated radiation in 1958 by observing the phenomenon of ionosphere absorption of electromagnetic waves [157]. American scholar Jurgen Schneider and former Soviet scholar A. G. Gapanov, V. V. Zheleznyakov independently proposed the concept of using the relativistic effect to make a spiral electron beam moving in a magnetic field interact with electromagnetic waves [158159]. Twiss and V. V. Zheleznyakov derived linear theory using a simple model. American scholar W. W. CHOW built a cyclotron backward oscillator in 1960 [160], which verified the feasibility of electron cyclotron resonance. The American scholar Hirshfield demonstrated the mechanism of electron cyclotron maser in experiments in 1964 [161]. The first gyrotron was invented in Gorky in the former Soviet Union (now Nizhny Novgorod, Russia) [18].

Driven by millimeter wave radar, electronic warfare, high-power microwave weapons, controlled nuclear fusion, high-energy physics and other applications, after decades of development, a series of electronic cyclotron masers (ECMs) (also known as cyclotron resonance masers (CRMs)) devices have been developed [162]. Especially, the gyrotron has been greatly developed. The gyrotron refers to a type of ECM device in which the interaction between the beam and the electromagnetic field occurs near the cutoff frequency of a uniform waveguide [162]. The simplest structure of a gyrotron is shown in Fig. 19. It consists of an electron gun (include the cathode and the anode), an interaction cavity, a collector, an output window, a magnet, etc.

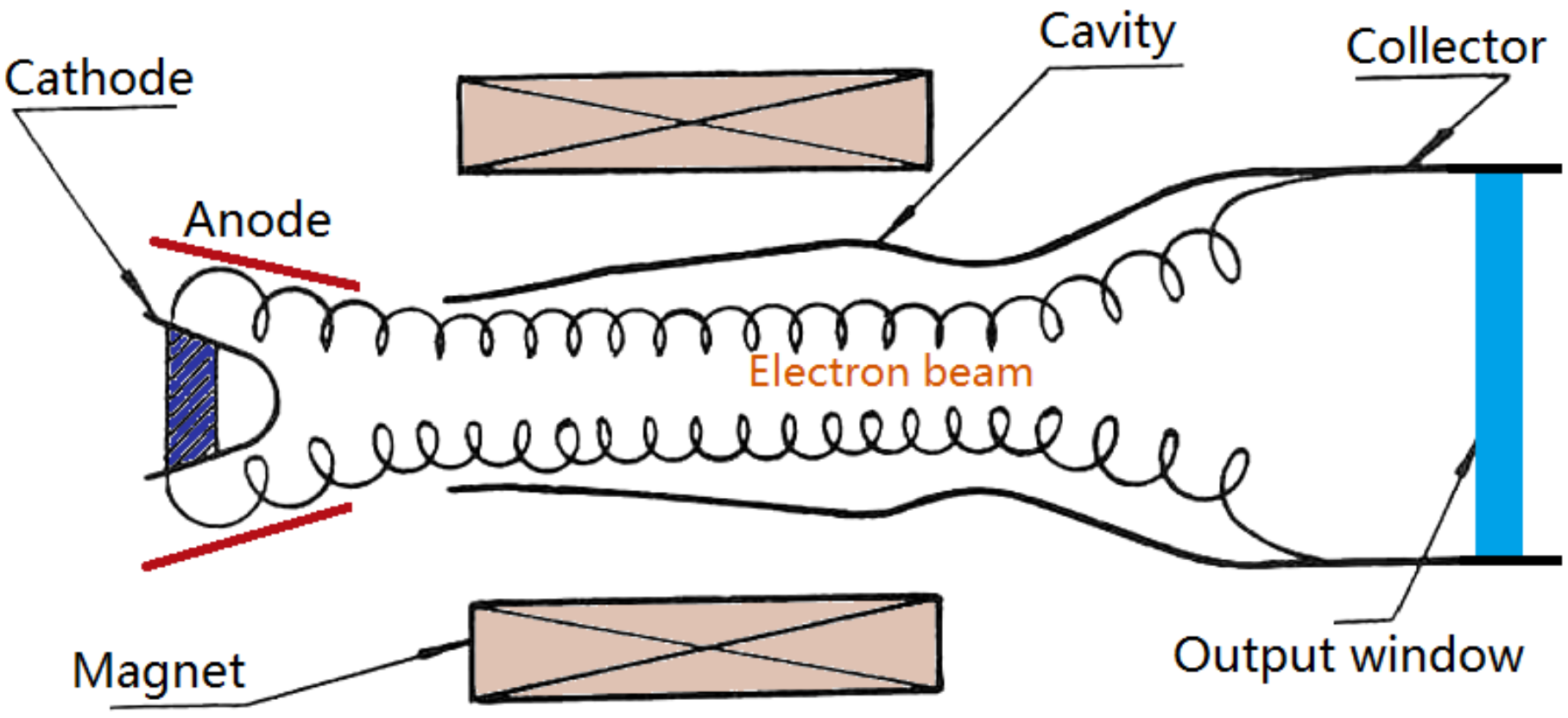

FIG. 19. The basic structure of the gyrotron.

After the electrons are emitted by the electron gun, they pass through a region where the magnetic field gradually increases, adiabatic compression occurs according to the following law, and the lateral velocity gradually increases.

$$\frac{p_{\perp}^{2}}{B} = \text{const.} \tag{8}$$

When in the following cyclotron resonance conditions, the electromagnetic waves can interact with the beam,

$$\omega - k_z v_z \approx n\Omega. \quad (9)$$

Where $\omega$ is the angular frequency of the electromagnetic wave, $k_z$ is the longitudinal wave number, $v_z$ is the electronic longitudinal speed, n is the harmonic order of the electron cyclotron resonance. After passing through the interaction cavity, the electrons pass through the area where the magnetic field gradually decreases, adiabatic decompression occurs, and finally enters the collector.

When electrons interact with waves, electromagnetic waves may cause changes in the axial velocity of the electrons and cause axial clustering; changes in the energy of the electrons may also cause changes in the electron cyclotron frequency, thereby causing angular clustering. From Eq. (9), it can be seen that when the electron beam interacts with the traveling wave, the change in the distribution of the axial velocity of the electron will cause a serious non-uniform Doppler broadening, which will cause the beam interaction efficiency to decrease. In order to solve this problem, the gyrotron usually chooses to work near the cut-off frequency with a small $k_z$, which greatly reduces the Doppler broadening, and the angular clustering plays a major role. Fig. 20 shows the interaction conditions of the waveguide mode dispersion curve and the electron cyclotron maser.

In addition to gyrotrons, a series of electron cyclotron maser devices have appeared, such as Gyro-TWT [163-165], Gyro-BWO [166], Gyro-Klystron, Gyro-Peniotron, Cyclotron AutoResonance Maser (CARM), Magnicon, etc. The gyro-TWT works in the traveling wave state, its operating point is far from the cutoff frequency, the electron beam and the waveguide mode remain synchronized in a relatively wide frequency range, and the device has a relatively wide bandwidth. The gyro-BWO operates under the condition that the longitudinal wave number is negative, thereby forming a backward wave oscillation. When the waveguide mode dispersion curve intersects with the electron beam cyclotron synchronization mode near the cutoff frequency, and $k_z$ approaches zero and is positive, it is the operating point of the gyro-klystron and gyro-monotron.

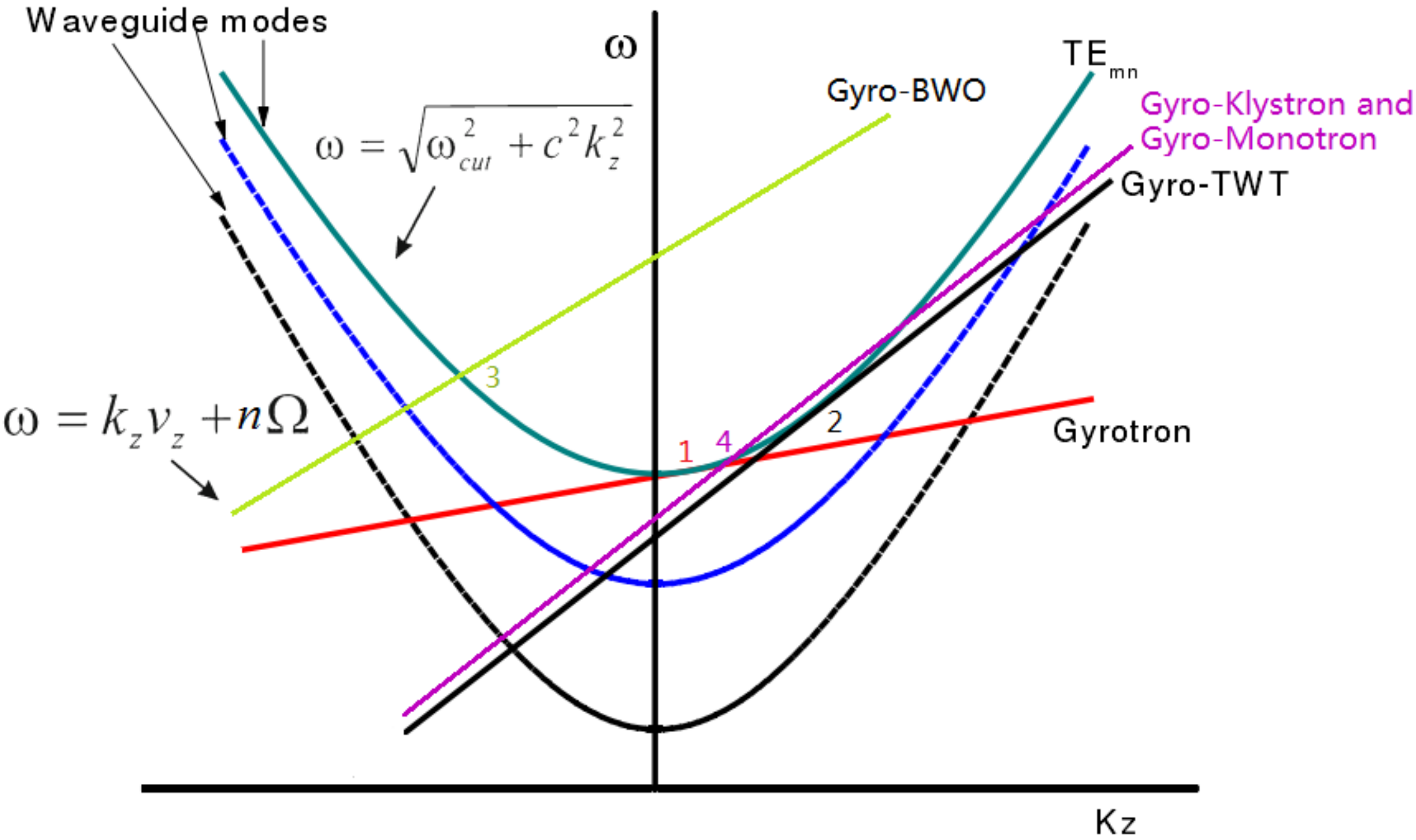

FIG. 20. The Brillouin diagram of ECMs. The curve in this figure is the waveguide mode dispersion curve, the cutoff frequency is $\omega_{cut}$. The red straight line is the working line of the gyrotron, the black straight line is the working line of the gyro-TWT, the green straight line is the working line of the gyro-BWO ($k_z$ is negative), and the purple straight line is the working line of the gyro-klystron and the gyro-monotron.

As shown in Fig. 21, there are two forms of electron beams formed by the electron gun of ECMs, that is, the large cyclotron electron beam and the small cyclotron electron beam. The latter is currently more commonly used. For the large cyclotron electron beams, the electron directly rotates around the center of the guide, and the electron beam interacts with the wave through angular clustering (relativistic) or the ***E***×***B*** translation effect (similar to the beam-wave interaction principle of the Peniotron, which is non-relativistic) [167,168]. For small cyclotron electron beams, the electrons emitted by the electron gun form a hollow electron beam, and each electron in the hollow electron beam is performing the same cyclotron motion. You can think of them as small circles one by one. In the same circle, there are many electrons rotating. They just have different positions on the circumference, that is, the electron beam interacts with the waves with different phases. The beam-wave interaction can be divided into three phases [162]: energy modulation; angular clustering; clustering deceleration. The revolving electron beam interacts with the high-frequency field excited in the high-frequency structure, thereby converting the lateral kinetic energy of the electron beam into electromagnetic wave energy in the required frequency band [169-172].

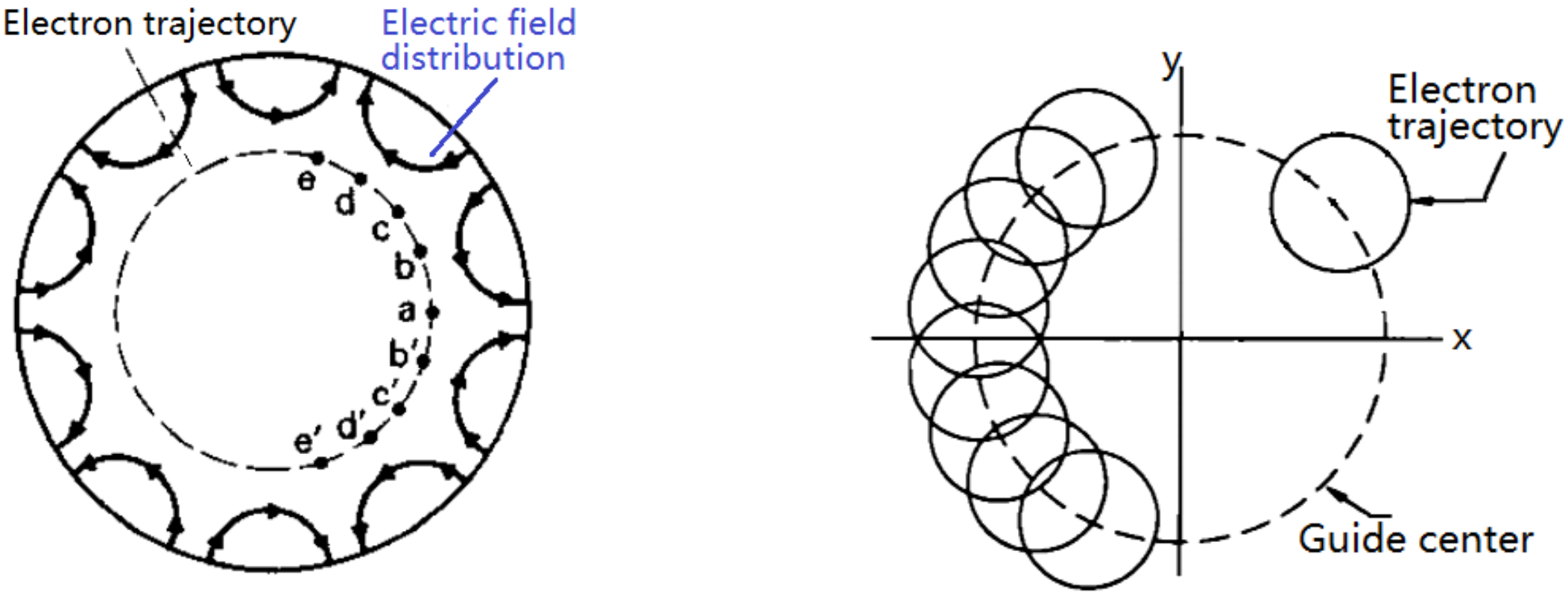

FIG. 21. The trajectories of the electrons of the large cyclotron electron beam and the small cyclotron electron beam. The harmonic number of the large cyclotron electron beam shown on the left is 5.

As shown in Fig. 19, the gyrotron consists of an electron gun, a magnet, a collector, an interaction cavity, and an output window. They are briefly described below.

An electron gun is a component that emits electrons. Vacuum devices such as gyrotrons and klystrons mostly use hot cathode electron guns. The types of hot cathode emission of electrons can be divided into space charge limited emission [173] or thermionic emission. The only criterion for measuring whether the electron supplied by the electron gun is limited by space charge or temperature is to look at the electric field strength on the cathode surface. If the electric field on the cathode surface is zero, or even negative (the appearance of a virtual cathode), it is in space charge limitation; If the electric field on the cathode surface is positive, the emission is limited by the temperature at this time, because as long as the cathode can emit electrons, it will be pulled away by the positive electric field on the surface.

Electron guns for non-relativistic magnetrons are mostly based on space charge limitation, and relativistic pulse magnetron electron guns are mostly based on temperature limitation [174]. Electron guns for O-type slow-wave electric vacuum devices such as klystron are mostly based on space charge limitation [175]. The gyrotron electron gun is a magnetron injection gun (MIG). Because the space charge affects the electron beam formation, increasing the space charge will increase the electron beam dispersion and reduce the output efficiency, so the gyrotron electron gun is mostly based on temperature limitation; There are some space charge limited electron guns designed for ECM devices [176,177].

As mentioned above, there are two types of gyrotron guns: the large cyclotron gun and the small cyclotron gun.

Most small cyclotron electron guns use MIGs. The hollow electron beams generated by a ring-shaped cathode generate a rotating motion under the action of an electrostatic magnetic field. Large cyclotron electron guns generally use a spherical ring-shaped cathode to generate hollow, non-rotating electron beams in the structure of the convergent Pierce electron gun. Then the Cusp magnetic field causes the electron beam to rotate around the central axis. The large cyclotron electron beam generated in this way enters the beam-wave interaction region, so that the electron beam excites the circularly polarized $TE_{m1}$ wave under the m-th cyclotron oscillation harmonic, and exchanges energy with electromagnetic waves.

The small cyclotron electron guns are widely used in the gyrotrons, gyro-TWTs, gyro-BWOs, gyro-klystrons, gyro-traveling-wave-klystrons, and many other devices; The large cyclotron electron guns are usually used in gyro-magnetrons, magnicons, Peniotrons [178], etc.

The small cyclotron electron gun was first proposed by Johnson and Dtskerson in 1964, and Gaponov of the former Soviet Union used it for gyrotron oscillators in 1965 [179]. The small cyclotron electron guns used in the gyrotrons have various forms, including a single anode electron gun in the form of a diode, a double anode electron gun in the form of a triode, an inverted electron gun, and a non-adiabatic gun. The former two are now more widely studied and applied.

Fig. 22 and Fig. 23 show the typical structure of the single anode electron gun and double anode electron gun. The double anode MIG introduces a modulation anode, which can more easily modulate the electron beam to generate an electron beam with a large working current, a relatively high aspect ratio (longitudinal velocity / lateral velocity), and a small velocity dispersion, but it increases the system complexity. The single anode MIG has a simple and compact structure, which is convenient for practical applications. However, under the condition of large working current, the generated electron beam has a relatively scattered speed, a small aspect ratio (longitudinal velocity / lateral velocity), and it is difficult to adjust.

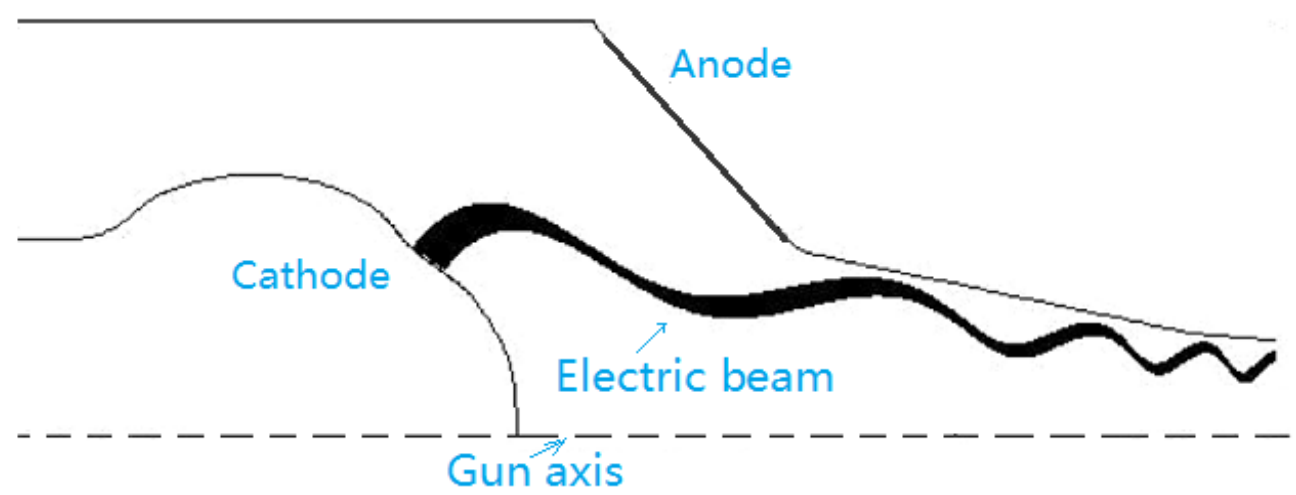

FIG. 22. The diagram of a typical single anode MIG.

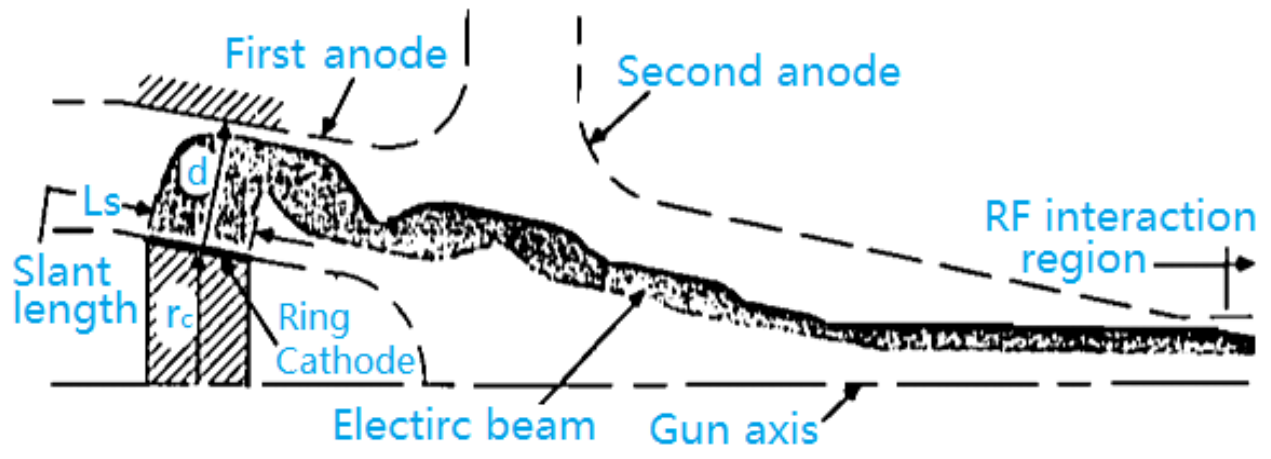

FIG. 23. The diagram of a typical double anode MIG.

The design of the gyrotron electron gun requires the selection of a suitable magnetic field distribution. The magnet is used to form the required magnetic field distribution in the electron gun, the interaction cavity, and the collector. It is generally required that the magnetic field is distributed symmetrically and uniformly in the interaction cavity, and the magnetic field intensity gradually decreases in the region of the collector and the electron gun.

The beam-wave interaction efficiency $\eta_{int}$ in the gyrotron is generally 30% ~ 50%, which means that after the beam-wave interaction zone, there is about 50% ~ 70% of the residual beam energy. The collector is used to absorb these residual electrons. The potential of the collector was the same as the interaction cavity in the earlier gyrotron design. In 1994, Sakamoto et al. first proposed a depressed collector [180]. The structure of the gyrotron using a single-stage depressed collector is shown in Fig. 24. The single-stage depressed collector is widely used in the high-power gyrotrons now.

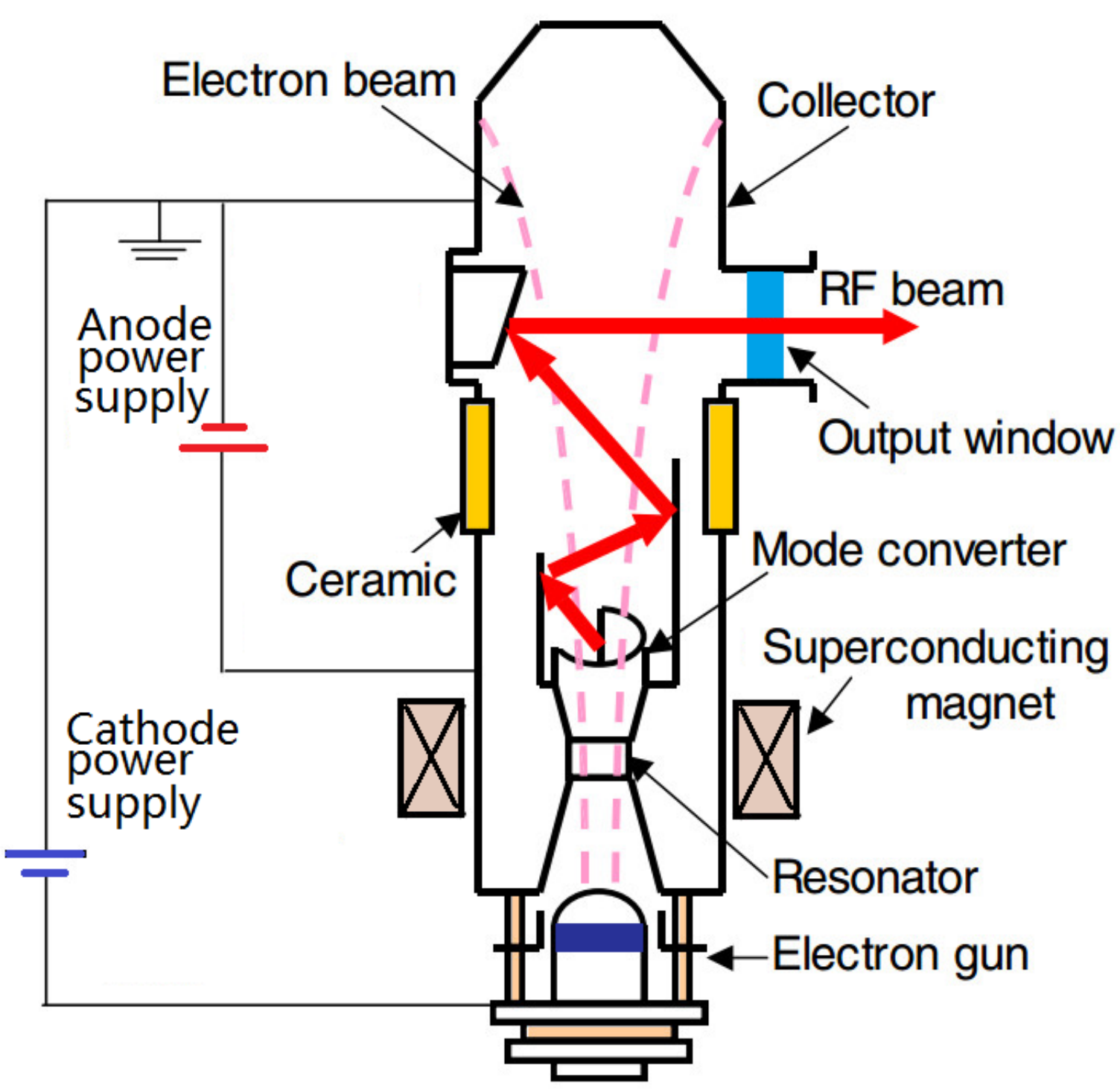

Fig. 24. The diagram of a gyrotron using a single-stage depressed collector.

The interaction between the electron beam and the microwave occurs in the interaction cavity. There are two forms of gyrotron interaction cavity, one is a cylindrical waveguide cavity, and the other is a coaxial cavity. The typical working modes in a cylindrical cavity are angularly symmetric $TE_{0n}$ mode, $TE_{mn}$ Whispering Gallery Mode (WGM) [167], and $TE_{mn}$ Asymmetric Volume Mode (AVM) [181]. The AVM refers to the mode of m >> 1, n> 2, and the characteristics of the cavity are between the $TE_{0n}$ mode and the WGM. Therefore, in the field of high-power long-pulse gyrotrons, the AVM are more widely used.

The gyrotron resonant cavity works in a higher-order mode. The waves in the higher-order mode have serious diffraction and loss and are not suitable for transmission. Therefore, they need to be converted into a low-loss low-order mode or a Gaussian beam. The "Vlasov" coupler structure is currently widely used [182]. The output mode is converted into an approximate Gaussian beam $TEM_{00}$ mode, and the microwave output direction is perpendicular to the electron beam direction.

The use of Vlasov structure benefits from the appearance of the CVD (chemical vapor deposition) diamond window. The output window is used to transmit RF waves while ensuring the vacuum inside the gyrotron. Materials that can be used as the output window include BeO, BN, Si3N4, SiC, sapphire, and CVD diamond. The CVD diamond material has strong compression resistance, small thermal expansion, high thermal conductivity, and low millimeter wave loss. It is currently the only one that can use the edge cooling method to pass 1 ~ 2MW continuous wave power. But the CVD diamond window is very expensive.

Due to the promotion of fusion [183] and other fields, the gyrotrons have been greatly developed in recent years, and is currently developing in the direction of high power and long pulses [184]. In order to achieve a long-pulse, high-power gyrotron, three main issues need to be considered: the heating of the cavity wall by microwwaves; the heating of the output window by microwaves; and the heating of the collector by electron beams and microwaves [185].

In order to solve the problem of high heat in the interaction cavity of long-pulse high-power gyrotrons, there are two methods. One is to increase the size of the cavity, but correspondingly requires a higher-order waveguide mode; the other is to improve the cooling effectiveness of the cavity. In order to solve the high heat problem of the output window, one method is to improve the window material; the second is to improve the mechanical design; the third is to choose the appropriate output mode. In order to solve the high heat problem of the collector, one method is to separate the wave from the electron beam as much as possible so that the wave cannot enter the region of the collector; the second method is to AC modulate the magnetic field of the collector so as to effectively scan the remaining electron beam along the collector wall; the third method is to develop the cooling technology to improve the cooling efficiency.

The progress of the ITER gyrotrons and the MW-level gyrotrons in recent years are shown in Table XIX and Table XX. Recently, at the Institute of Plasma Physics Chinese Academy of Sciences (ASIPP), we successfully realized the 3000s operation of a new GYCOM gyrotron with an RF output power of 700 kW. The development of ECMs such as gyrotrons, gyro-TWTs, gyro-BWOs, gyro-klystrons, gyro-Peniotrons, cyclotron autoresonance masers (CARMs), and magnicons can refer to reference [186].

TABLE XIX. The research progress of ITER gyrotrons. In this table, among the parameters of JAEA's multi-frequency gyrotron, the power and efficiency parameters are corresponding to the corresponding frequency, separated by a left slash, where the efficiency is measured in the continuous short pulse mode; In the GYCOM gyrotron parameters, the maximum pulse length is different at different power outputs, and they are listed separately and separated by a left slash. JAEA is short for Japanese Atomic Energy Agency; GYCOM is a Russian gyrotron producer; the European Gyrotron Consortium (EYGC) consists of Swiss CRPP, German KIT, French TED (Thales Electronic Device), Greek HELLAS and Italian CNR / ENEA, etc.

| Institute | Model | Freq. [GHz] | Mode | Power [kW] | Efficiency [%] | Pulse length [s] | Cavity structure | Electron gun form |
|---|---|---|---|---|---|---|---|---|
| JAEA[187] | | 170/137/104 (Multi-Freq.) | TE31,11/ TE25,9/ TE19,7 | 905/540/ Untested | 27.3/23.0/ 23.7 | 75/20/ Short pulse | Cylinder | Triode |
| JAEA | | 170 | TE31,8 | 1000 | 55 | 800 | Cylinder | Triode |
| GYCOM/IAP[188 189] | V-10 | 170 | TE25,10 | 1000<br>960<br>800<br>750 | 54 | 570<br>578<br>800<br>1000 | Cylinder | Diode |
| GYCOM/IAP | V-11 | 170 | TE25,10 | 960<br>850 | 54 | 400<br>1000 | Cylinder | Diode |
| GYCOM/IAP[189] | | 170 | TE28,12 | 2000 | 34 | $10^{-4}$ | Cylinder | Diode |
| EGYC | | 165 | TE31,17 | 2200 | 48 | 0.001 | Coaxial cavity | Diode |
| EGYC | | 170 | TE34,19 | 2200 | 30 | 0.001 | Coaxial cavity | Diode |

TABLE XX. The research progress of gyrotrons with $TEM_{00}$ mode, MW-level output power [190].

| Institute | Freq. [GHz] | Mode | Power [MW] | Efficiency [%] | Pulse length [s] |
|---|---|---|---|---|---|

| CPI/MIT/GA | 110 | $TE_{22,6}$ | 1.5 | 50 (SDC) | $10^{-6}$ |
|---|---|---|---|---|---|
| | | | 1.2 | 41 (SDC) | $5*10^{-3}$ |
| TOSHIBA/JAEA | 110 | $TE_{22,6}$ | 1.5 | 45 (SDC) | 4.0 |
| | | | 1.4 | 45 (SDC) | 9.0 |
| CPI[191] | 140 | $TE_{28,7}$ | 0.9 | 35 (SDC) | 1800 |
| TED/KIT/CRPP[192] | 140 | $TE_{28,8}$ | 0.92 | 44 (SDC) | 1800 |
| GYCOM/IAP[193] | 140 | $TE_{22,8}$ | 0.9/0.83/0.56 | 50 (SDC) | 10/95/1000 |
| GYCOM/IAP | 140 | $TE_{22,8}$ | 0.7 | 45 (SDC) | 3000 |
| TOSHIBA/JAEA | 154 | $TE_{28,8}$ | 1.2 | 27 | 1.0 |
| KIT/EFDA | 165 | $TE_{31,17}$ (coax.) | 2.2 | 48 (SDC) | $10^{-3}$ |
| GYCOM/IAP | 170 | $TE_{28,12}$ | 1.5 | 49 (SDC) | 2.5 |
| | | $TE_{25,10}$ | 1.2 | 52 (SDC) | 100 |
| TED/EGYC/F4E | 170 | $TE_{34,19}$ (coax.) | 2.1 | 46 (SDC) | $10^{-3}$ |
| TOSHIBA/JAEA | 170 | $TE_{31,11}$ | 1.4 | 28 | $10^{-3}$ |
| | | | 1.1 | 45 (SDC) | 5.0 |

At present, gyrotrons are also starting to develop towards multiple operating frequencies. There are various gyrotron frequency tuning methods [194], including mechanical tuning [195], electrical tuning (adjusting the acceleration voltage or modulation voltage [196,197]), magnetic field strength tuning [198] (adjust the magnetic field strength by adjusting the magnetic field coil current), high-harmonic adjustment [199], and so on. JAEA, EYGC, GYCOM, etc. have developed the dual-band or multi-band gyrotrons. For example, KIT has designed a multi-frequency tuned gyrotron. It tested 3 ms pulses without depressed collector [200], and can output 1 MW power at frequencies such as 127.1GHz, 127.4GHz, 162.5GHz, and the efficiency is about 25%. There are three main challenges for dual-frequency or multi-frequency gyrotrons [201]. The first is the output window. The thickness of the output window should be an integer multiple of the half wavelength of the diamond. The second is a cavity. The cutoff frequency of the cavity must be close to all operating frequencies. The third is a mode converter. The mode converter must be compatible with all operating frequencies. The same radiation angle for all operating modes is very important.

We believe that the development of gyrotrons will become better and better, and continue to develop in the direction of higher power, higher frequency, longer pulse width, frequency tuning, higher stability, and easier use. It is believed that in the future, users can use equivalent circuit models to characterize gyrotron parameters [202], and develop gyrotron systems like chips.

## 4.2. Free Electron Laser

Free electron lasers (FELs) can output waves range in frequency from microwave to ultraviolet light. In the case of a spatially periodic magnetic or electric field (undulator/wiggler),

$$\Omega_b = k_w v_z = \frac{2\pi}{\lambda_w} v_z. \tag{10}$$

Where $\Omega_b$ is the transverse oscillation frequency (bounce frequency) of the moving charges; $v_z$ is the electron beam velocity; $\lambda_w$ is the wiggler field spatial period.

If the electron beam is highly relativistic ($v_z \cong c$), the radiation will have a much shorter wavelength,

$$\lambda \cong \lambda_w / 2\gamma^2, \tag{11}$$

where $\gamma$ is the relativistic Lorentz factor. The FELs use relativistic beam technology to greatly increase the frequency of output electromagnetic waves,

$$\omega = \frac{2\pi}{\lambda} v_z \cong 2\gamma^2 \Omega_b. \tag{12}$$

The basic FEM configuration is shown in Fig. 25. The electrons in an injected electron beam undulate in the periodic magnetic field of the wiggler. The magnetostatic wiggler is the most common type, but an electrostatic wiggler or the oscillatory field of a strong electromagnetic wave can also work.

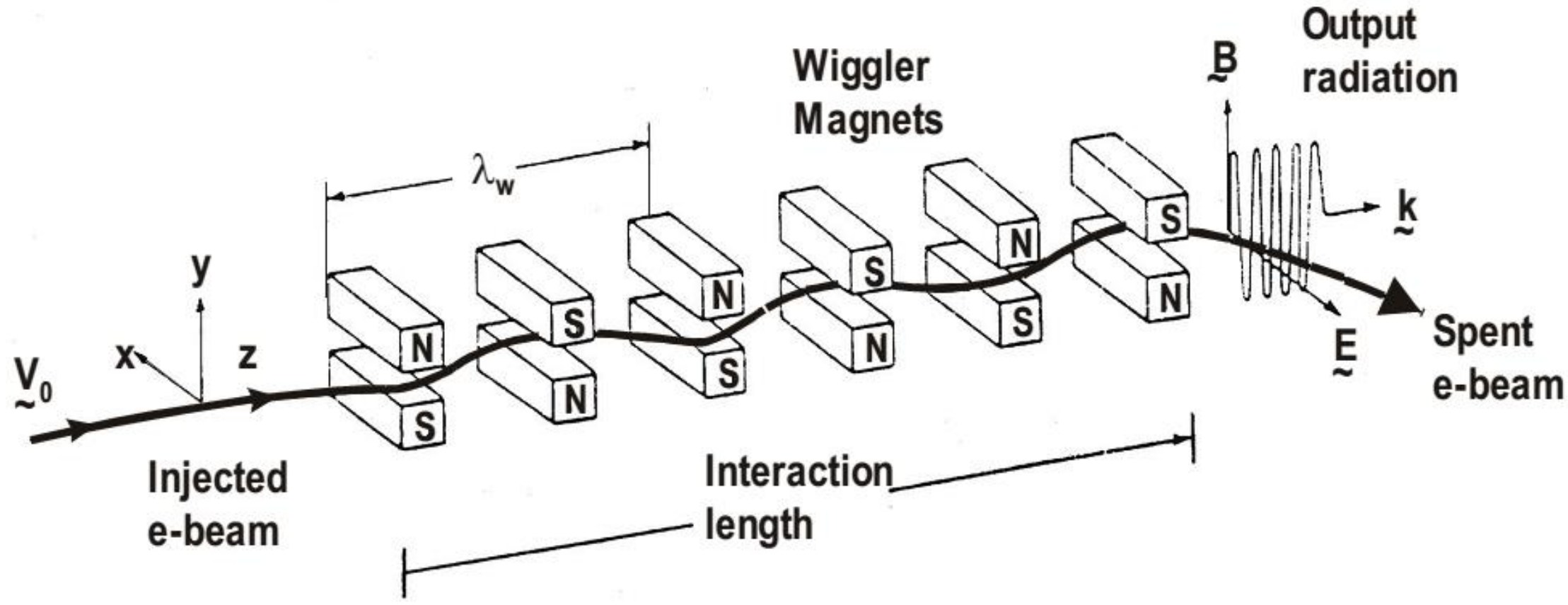

FIG. 25. The basic FEM configuration.

We call the free electron laser with an output wavelength greater than 0.5 mm as FEM (free electron maser). The maximum output wavelength of the FEM is 1.1 m employing a 760 V, 0.1 A electron beam ($f$ = 275 MHz, $\lambda_w$ = 0.04 m, $B_w$ = 0.04 T; 3 W, 2 ms pulses) [203]. Some important advances in free electron lasers are shown in XXI. State-of-the-art of millimeter and submillimeter wave FEMs and the comparison of gyrotrons and FEMs for ECRH for magnetic confinement fusion can be found in Ref. [186].

TABLE XXI. The development status of FELs.

| **Band** | **Peak Power [W]** | **Average power [W]** | **Freq. [Hz]** | **Effiency [%]** | **Gain [dB]** | **Pulse [s]** | $B_w$ **[T]** | $\lambda_w$ **[m]** | **Voltage [V]** | **Curretn [A]** | **Institution** | **Type** |
|---|---|---|---|---|---|---|---|---|---|---|---|---|
| P | 3 | | 275M | 3.9 | | 2m | 0.04 | 0.04 | ~0.76k | ~0.1 | TAU | Oscillator |
| C[204] | | 175.8 | ~7.18G | 10.2 | | CW | 0.044 | 0.038 | 129.1k | 13.3m | UL, UK | Oscillator |
| X | | 20 | ~8G | 1.2 | ~13 | CW | 0.044 | 0.038 | 129.1k | 13.3m | UL, UK | Amplifier |
| Ka[205] | 15M | | 30.2G | | | 120n | | | | | IAP; JINR | Oscillator |
| THz[206,207] | Up to 1M | Upt to 0.5k | 1.25T-60T | ~0.5 | | Up to 100p | | | | | BINP | Oscillator |

| | | | | | | | | | |
|---|---|---|---|---|---|---|---|---|---|
| THz[208] | 167M | 235 | 10T | | 300f | | | TJNAF | Oscillator |
| IR[209,210] | | 1.72k | 96.8T | 0.8 | CW | 47.8M | 4.4m | TJNAF | Oscillator |
| IR | 417M | 2.3k | 60T-300T | | 300f | | | TJNAF | Oscillator |
| IR[211,212] | | Up to 10k | 21.4T-300 T | | | | | TJNAF | Oscillator |
| ~Light | | Up to 1k | 0.3P-1.2P | | | | | TJNAF | Oscillator |
| ~Light | 26.7M | 75.2 | 0.33P-0.8 1P- | | 300f | | | TJNAF | Oscillator |

Note: TAU is short for Tel-Aviv University, Israel. UL, UK is short for University of Liverpool, UK. JINR is short for Joint Institute for Nuclear Research, Russia. BINP is short for Budker Institute of Nuclear Physics, Russia. TJNAF is short for Thomas Jefferson National Accelerator Facility, United States.

# 5. Vircator (VIRtual CAthode oscillaTOR)

As a kind of high-power microwave device with a strong non-linear space charge effect, the virtual cathode oscillator generally does not require an external guidance magnetic field. Compared with other types of high-power microwave sources, it has the advantages of simple structure, low requirements for the quality of the electron beam, high power capacity, relatively easy tuning and low impedance. However, it also has shortcomings such as relatively low beam-wave conversion efficiency, a cluttered frequency spectrum, and impure modes. The Vircator has a GW-level output capability in the frequency range of 0.4 GHz to 17 GHz. In the 1990s, it flourished in the United States and Russia, and there have been few related reports recently.

The basic vircator configuration is shown in FIG. 26. The electrons are attracted to a thin anode, such as an aluminized plastic film that is connected to the ground. Due to the intensity of the electron beam, many electrons enter the region outside the anode through the anode, thereby forming a virtual cathode. The electron beam must be strong enough to exceed the space charge limiting current in this region, causing microwave oscillations.

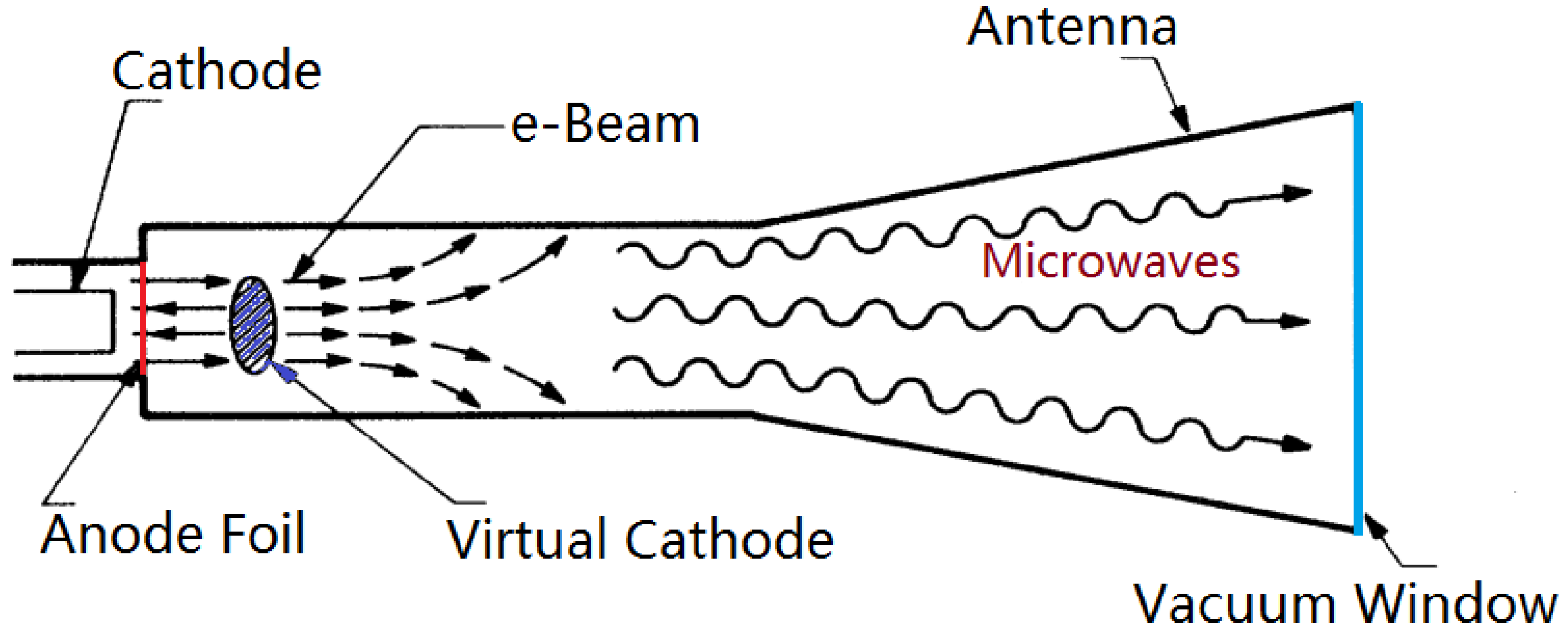

FIG. 26. The basic vircator configuration.

The central frequency follows the beam plasma frequency. It is tuned by varying the current density with anode-cathode (A-K) gap adjustments. Variation of the cathode radius is shown to have little effect on the oscillation frequency, however, changing the anode-cathode gap is shown to have a strong effect on the frequency.

There are multiple forms of Vircators, include Reditron (Reflected Electron DIscrimination oscillator) [213],

side-shooting virtual cathode oscillator (SSVCO) [214], vircator with electron beam pre-modulation using two anode foils (VEBP) [215], etc. The research progress is shown in Table XXII.

TABLE XXII. The development status of the Vircators.

| **Band** | **Peak Power [W]** | **Average power [W]** | **Freq. [Hz]** | **Effiency [%]** | **Pulse [s]** | **Voltage [V]** | **Current [A]** | **Mode** | **Institution** |
|---|---|---|---|---|---|---|---|---|---|
| P[216] | 22G | 7.7G | 0.78G | ~0.5 | 24.3n | 6.7M | 225k | $TM_{0n}$ | HDL, etc. |
| P-C[217] | 200M-500M | | 0.45G-5.5G | ~0.5 | ~100n | 600k-800k | 50k-120k | $TM_{01}$ | PI |
| L[218] | 7.5G | | 1.17G | 2.3 | 10n | 4M | 80k | $TM_{01}$ | AFWL |
| S[215] (VEBP) | 1G | | 2.1G | ~5 | 50n | 1M | 20k | $TE_{10}$ | IHCE |
| S[219] (VEBP) | 300M | | 2.2G | ~5 | 130n | 300k | 20k | $TE_{11}$ | IHCE |
| S[213] (Reditron) | 1.6G | | 2.46G | ~6 | 30n | 1.3M | 20k | $TM_{01}$ | LANL |
| C | 300M | | 4.41G-4.69G | ~3 | 10n | 400k-480k | 20k-25k | $TM_{01}$ | NUDT |
| C[220] | 4G | | 6.5G | 3.3 | 10n | 1.5M | 80k | $TM_{01}$ | LLNL |
| Ku[9] | 500M | | 17G | 0.3 | 12n | 1.8M | 90k | $TM_{03}$ | LANL |
| Ka | 100M | | 30G-40G | 0.06 | 12n | 1.8M | 90k | $TM_{03}$ | LANL |

Note: HDL is short for Harry Diamond Laboratories, USA. AFWL is short for Air Force Weapons Laboratory, USA. LANL is short for Los Alamos National Laboratory, USA. LLNL is short for Lawrence Livermore National Laboratory, USA.

# 6. Summary

The nature of all vacuum tube microwave sources utilizes the interaction of electrons and waves. There are also some new beam-wave interaction structures that have been designed and studied [221]. There are also studies that use multiple sources to combine into one high-power source [6,221].

We have calculated the experimental results of various vacuum microwave tubes. Considering three indicators of frequency, power and pulse width, which are shown in Fig. 27-30. As can be seen from the figures, power, frequency, and pulse width are contradictory parameters. The larger the output power, the lower the frequency and the smaller the pulse width.

FELs have the ability to output larger frequencies than other vacuum tubes. TWTs, BWOs, SWOs, MWCGs, RDGs, Vircators, etc. can produce GW-level output power in the 10 GHz band, but the pulses are less than 1 μs.

Vacuum tubes that can achieve continuous wave operation include TWTs, Klystrons, FELs, and Magnetrons, with continuous wave output power up to 1MW.

Vacuum tubes that can generate frequencies of the order of 100 GHz and above include Klystrons, TWTs, Gyrotrons, BWOs, Magnetrons, SWOs, FELs, Orotrons, etc. The output power of SWOs can reach 10 MW level, but the pulse width is only ns level. The Gyrotrons output power at the MW level, the maximum pulse width can reach 3000s. In addition, The TWTs (folding TWTs) can output an average power of 100 W at 91.4 GHz and 39.5 W at 233 GHz. The Klystrons (EIKs) can output a peak power of 2 kW (average power 28 W) at 94.05 GHz and an average power of more than 9 W at 220 GHz. The Magnetrons (space harmonic magnetrons) can output a peak power of up to 20 kW at 95 GHz and a peak power of 1.7 kW at 225 GHz. The Orotrons can output a power of 50 mW at 140 GHz and 30 mW at 370 GHz.

The maximum power density record was created by FELs. The average power density of the FELs can reach $Pf^2 = 1.72\times10^{-3}\times(96.8\times10^3)^2 \approx 1.6*10^7$ [MW][GHz]$^2$ in 2000s, and $Pf^2 = 2.3\times10^{-3}\times(3\times10^5)^2 \approx 2.1*10^8$ [MW][GHz]$^2$ in 2010s. The development of vacuum devices still follows average power density Moore's Law now.

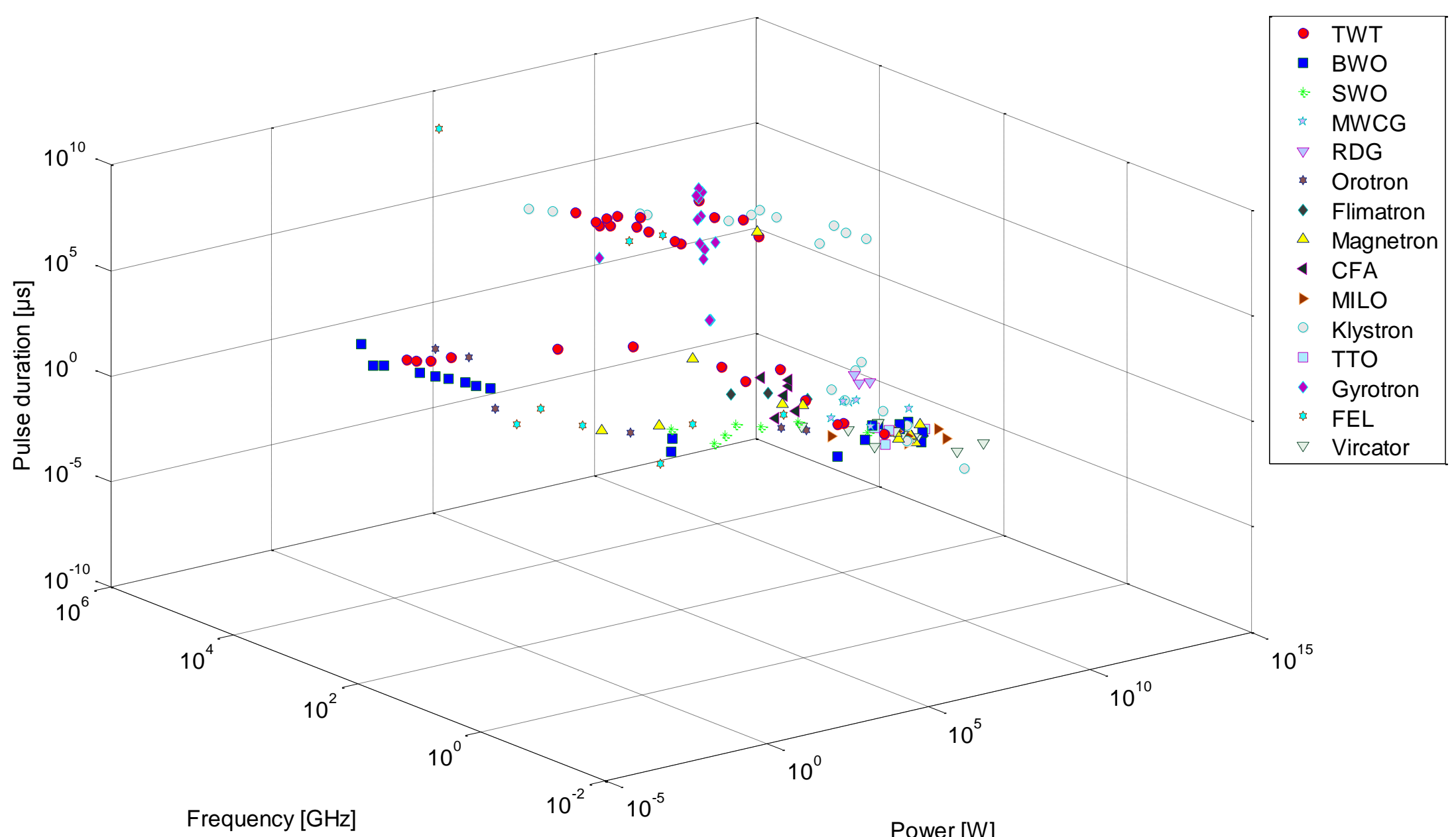

FIG. 27. The ‘frequency – power - pulse duration’ experimental results of various vacuum microwave tubes.

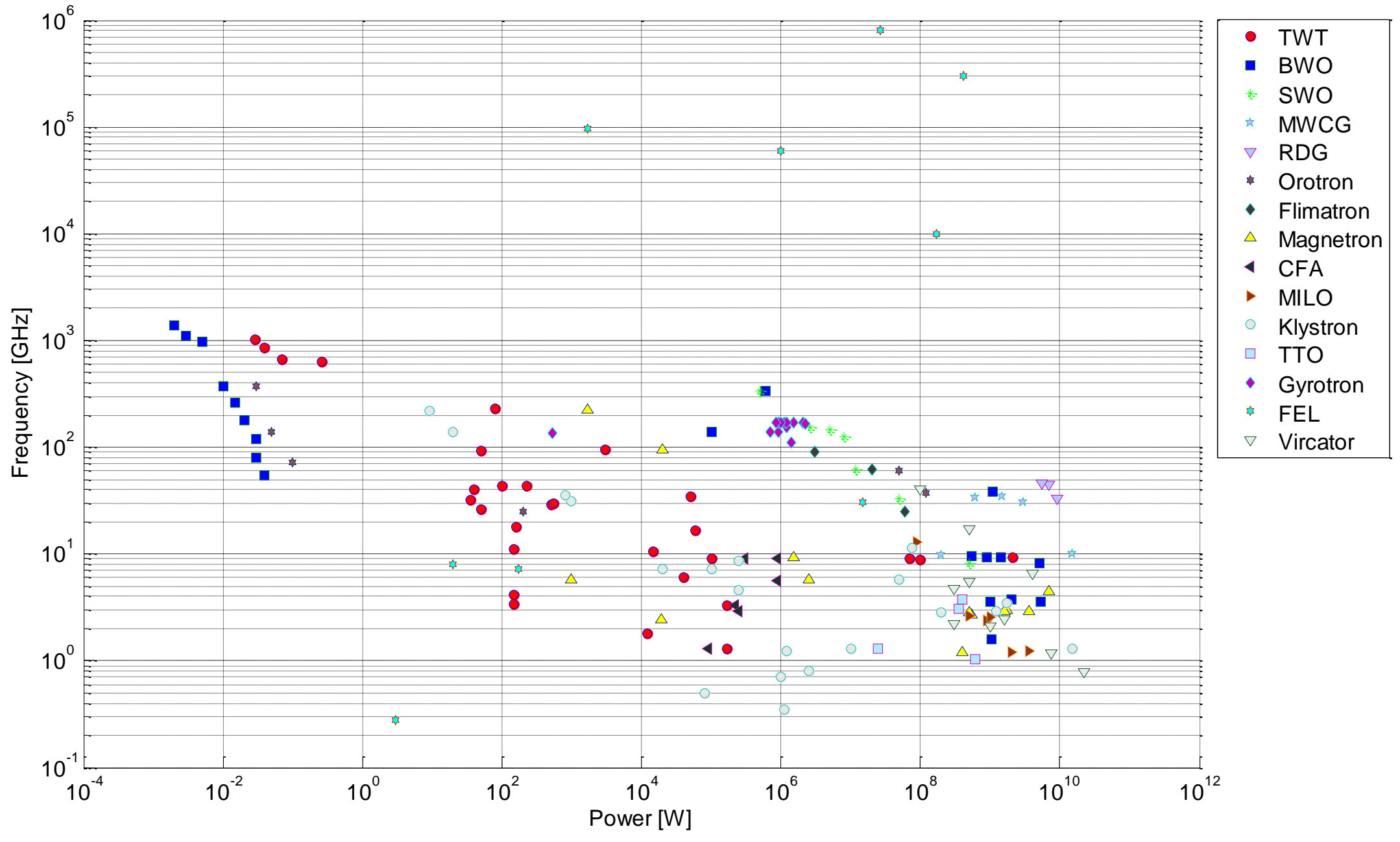

FIG. 28. The ‘frequency - power’ experimental results of various vacuum microwave tubes.

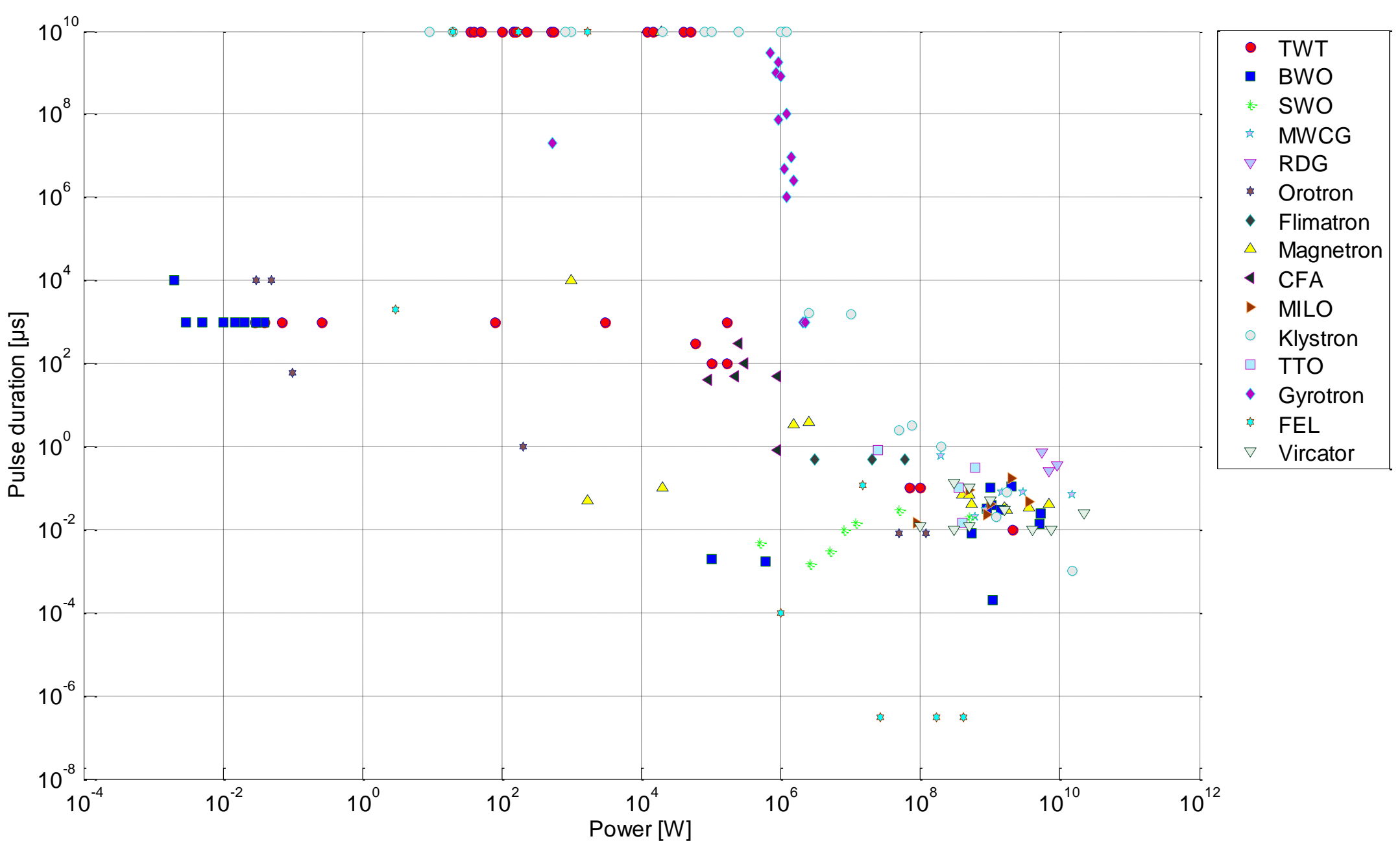

FIG. 29. The 'pulse duration - power' experimental results of various vacuum microwave tubes.

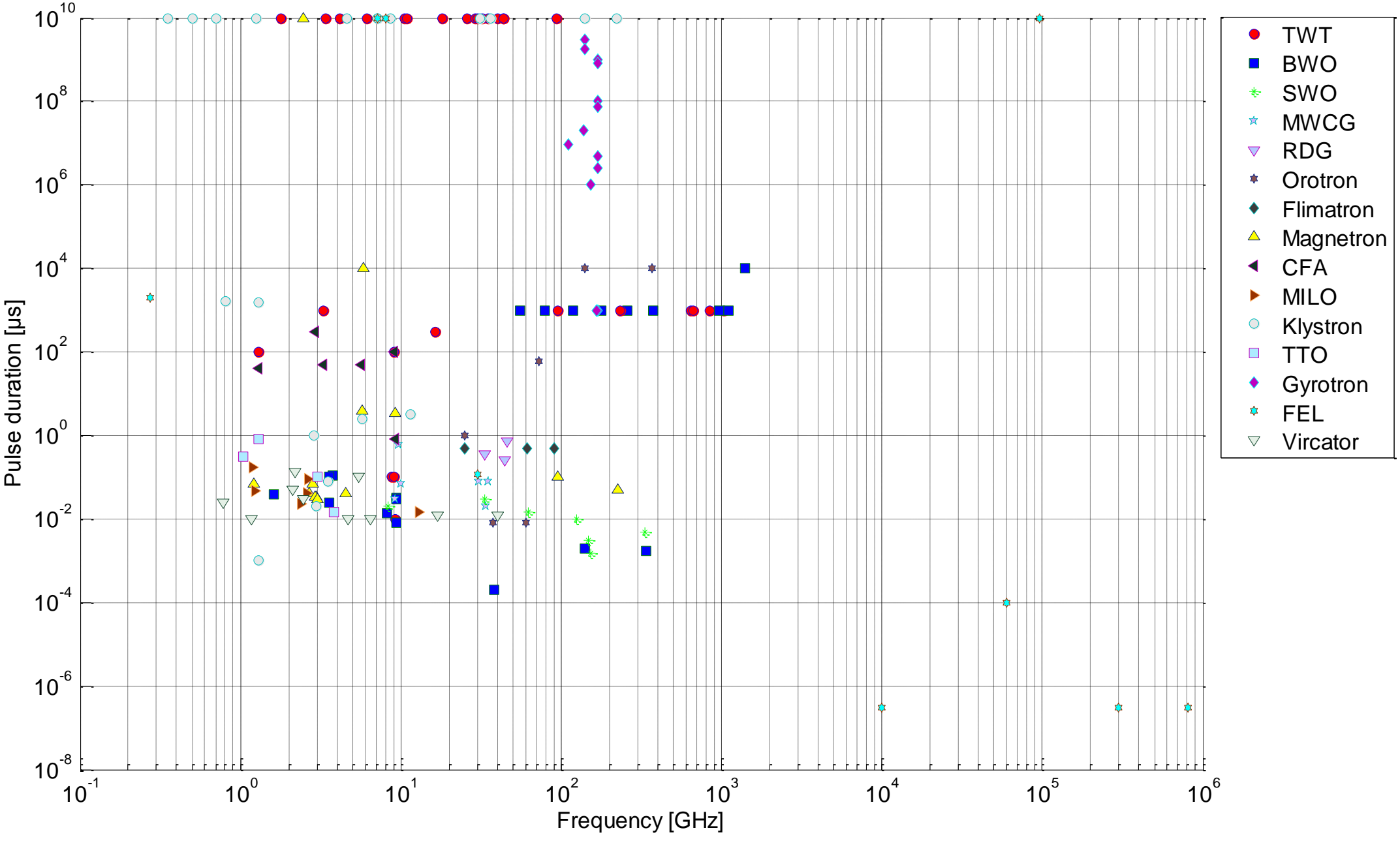

FIG. 30. The 'pulse duration - frequency' experimental results of various vacuum microwave tubes.

## Acknowledgments

This work was supported in part by the National Key R&D Program of China under Grant No.

2017YFE0300401 and 2016YFA0400600, and the National Magnetic Confinement Fusion Science Program of China under Grant No. 2015GB102003 and 2018YFE0305100.